%% file: main.tex
\documentclass[sigconf]{acmart}

\usepackage{algorithm}
\usepackage[noend]{algpseudocode}
\usepackage{multirow,multicol,xspace}

\newif\ifcountdistinct

\ifdefined\fullversion
    \countdistincttrue
\else
    
    \countdistinctfalse
\fi

\usepackage{etoolbox}

\ifcountdistinct
	\newcommand{\countdistinct}[1]{#1}
	\newcommand{\countdistinctapp}[1]{}
\else
    \newcommand{\countdistinct}[1]{}
    \newcommand{\countdistinctapp}[1]{#1}
    
\fi

\newif\ifcomm
\commtrue

\ifcomm
	\newcommand{\mycomm}[3]{{\footnotesize{{\color{#2} \textbf{[#1: #3]}}}}}
\else
    \newcommand{\mycomm}[3]{}
\fi

\usepackage{listings}
\definecolor{codegray}{gray}{0.9}

\lstdefinestyle{sql}{
    language=SQL,
    backgroundcolor=\color{codegray}, 
    breaklines=true,                  
    captionpos=b,                     
    frame=single                      
}

\usepackage{wrapfig}

\usepackage{graphicx}
\usepackage{comment}

\usepackage{caption}
\DeclareCaptionFont{large}{\normalfont}  
\usepackage{amsthm}
\usepackage{amsmath}
\usepackage{algorithm}
\usepackage{subfig}
\usepackage{xcolor}
\usepackage{thm-restate}
\usepackage{cleveref}

\input{macros}

\title{Counter Pools: Counter Representation
For Efficient Stream Processing}

\newtheorem*{observation*}{Observation}

\begin{document}
\title{Counter Pools: Counter Representation for Efficient~Stream~Processing}

\author{Ran Ben Basat}
\affiliation{%
  \institution{UCL}
  \city{}
  \country{}
}

\author{Gil Einziger}
\affiliation{%
  \institution{BGU}
  \city{}
  \country{}
}

\author{Bilal Tyah}
\affiliation{%
  \institution{BGU}
  \city{}
  \country{}
}

\author{Shay Vargaftik}
\affiliation{%
  \institution{VMware Research}
  \city{}
  \country{}
}

\settopmatter{printacmref=false} 
\renewcommand\footnotetextcopyrightpermission[1]{} 
\pagestyle{plain} 

\begin{abstract}
%
Due to the large data volume and number of distinct elements, space is often the bottleneck of many stream processing systems. The data structures used by these systems often consist of counters whose optimization yields significant memory savings. 
The challenge lies in balancing the size of the counters: too small, and they overflow; too large, and memory capacity limits their number.

In this work, we suggest an efficient encoding scheme that sizes each counter according to its needs. Our approach uses fixed-sized pools of memory (e.g., a single memory word or 64 bits), where each pool manages a small number of counters.  We pay special attention to performance and demonstrate considerable improvements for various streaming algorithms and workload characteristics.
\end{abstract}

\maketitle

\begin{figure*}[]
        \subfloat[Fraction of counters that \emph{do not} fit in a given \mbox{number~of~bits.}]{\label{fig:CCDF}
        \includegraphics[width =0.4121058727893405\linewidth]{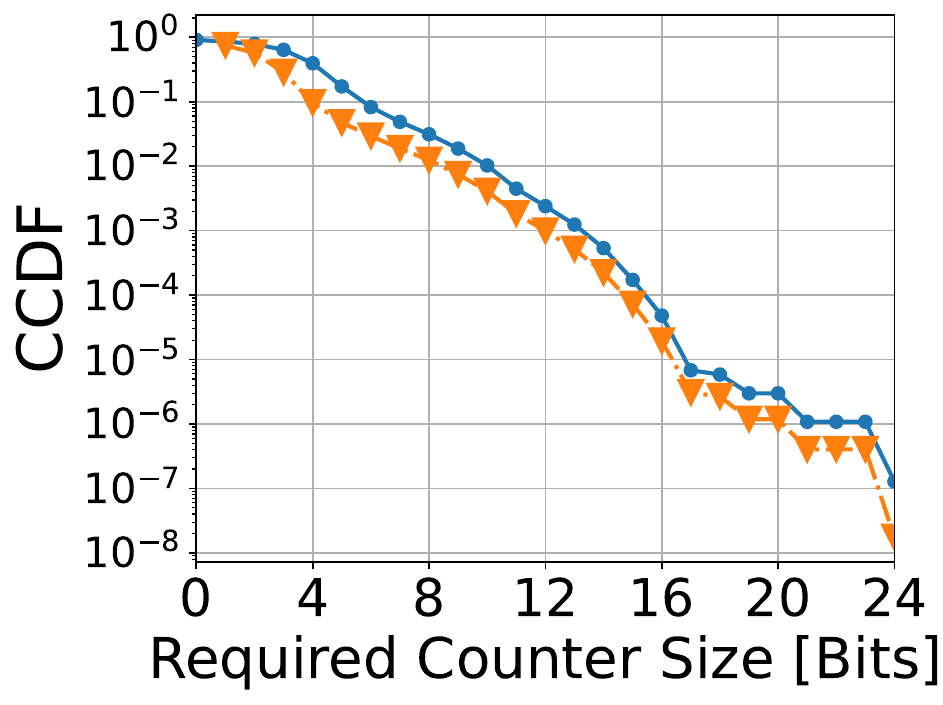}}
        \centering
        \hspace{12mm}
        \subfloat[\mbox{Fraction of counters that fit in a given number of bits.}]{\label{fig:CDF}
        \includegraphics[width =0.4121058727893405\linewidth]{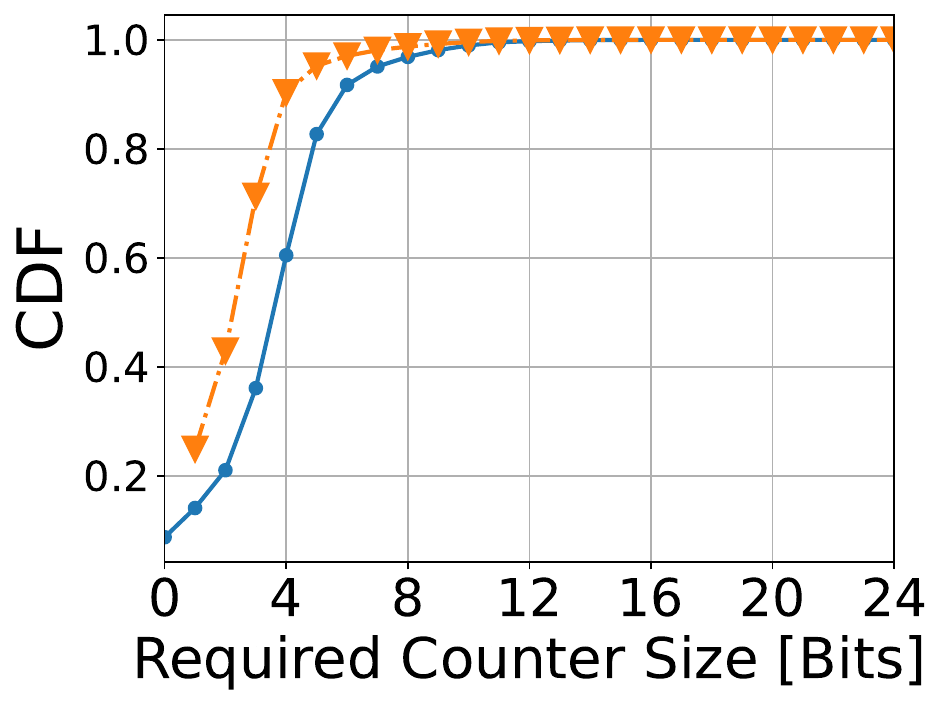}}    \\
        \centering
        \vspace{3mm}
        \includegraphics[width =0.493405\linewidth]{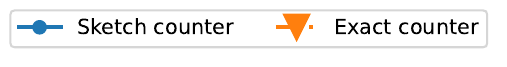} 
        \caption{Distribution of required counter sizes for the flows in the NYC 2018~\cite{CAIDA2018} workload for one exact counter per flow and a 2MB count min sketch. Notice that the sketch counters are slightly bigger due to hash collisions, but the general trend is very similar. In both cases, there is potential for memory savings by a counter-resizing scheme.}\label{fig: CounterSizeCmpNYC}
\end{figure*}





\section{Introduction}
Counter arrays are at the core of network measurement and stream processing systems and are a fundamental building block for various applications. These include sketches~\cite{CountMinSketch,ABF,PyramidSketch,AEE,shahout2023together}, traffic histograms, table-based network  measurements~\cite{SpaceSavings,SpaceSavingIsTheBest,RAP}, natural language processing~\cite{NLPsketches}, load balancing in networks~\cite{LoadBalancing}, and forensic analysis~\cite{Forensic2}.

\looseness=-1
As the data volume and number of distinct elements increase, the size of such arrays is often constrained by the available space.
As a consequence, algorithms trade off space for accuracy in approximate algorithms (e.g., sketches) and speed in exact ones (e.g., hash tables).
For example, approximation tasks include measuring per item frequency (or frequency histogram)~\cite{ConextPaper}, item distribution entropy~\cite{DBLP:conf/sigmetrics/LallSOXZ06}, and top-$k$/heavy hitters~\cite{RAP,HeavyHitters}, all attempting to optimize the space-accuracy tradeoff. 
Similarly, hash tables with lower load factor (i.e., the ratio between the number of used and allocated entries) tend to offer better insertion and query speeds.
Therefore, there is a significant interest in optimizing the counter arrays' space requirements by optimizing their representation.

The first choice, when it comes to counter representation, is to decide on the size (in bits) of each counter. For example, it is common to use 32-bit counters in many sketch implementations (e.g., see~\cite{cormode2008finding}) to ensure fast access for read and updates while avoiding overflows.
Intuitively, optimizing counter sizes is challenging. If we size counters for the maximal possible value, we waste space, as data skew often means that the average counter is significantly smaller.
On \mbox{the other hand, if we size them too small, they could overflow. }

As a motivating example, consider the flow sizes in a measurement segment of the NYC 2018 workload~\cite{CAIDA2018} containing 2.5M elements (network connections, also known as flows) and 98M packets. In Figure~\ref{fig:CCDF}, we illustrate the flow sizes in the measurement and the sketch counter size for a Count-Min-Sketch~\cite{CountMinSketch} of size 2MB. Here, `Exact Counter' refers to allocating a counter for each flow for an exact size measurement.
Notice that the largest flow requires 24 bits, and thus, in a fixed-sized counter architecture, we need at least 24 bits per counter. For the sketch the situation is worse, as the largest counters may require even more bits to avoid overflows in case multiple large flows have hash collisions and the aggregated value exceeds $2^{25}$. Moreover, it is hard to determine the correct size ahead of time without knowing the flow size distribution, which is only known after the measurement.


At the same time, as shown in Figure~\ref{fig:CDF}, the overwhelming majority of flows require counters utilizing less than 7 bits. This observation is valid not only for flow sizes but also for shared sketch counters. Hence, there is a potential opportunity for memory savings if we can engineer a dynamic counter-sizing scheme that strikes a balance between the speed required for stream processing, such as networking workloads, and the space efficiency to exploit these differences while accounting for potential overflows.

The consequences of overflows depend on the application. For approximation algorithms, one may be able to account for the overflows as part of the overall error and still provide provable guarantees at reduced space.
A recent approach suggests handling overflows by merging the counters~\cite{SALSA,gong2017abc}. This approach solves the overflow problem but increases error due to hash collisions on the overflowing counters. 
Moreover, such counters often belong to the heavy-hitter items, which are especially important to estimate precisely.  
Also, such an approach does not work when we do not allow any error and require exact representation. 

This work proposes a practical dynamic resizing approach based on fixed-sized pools of memory (e.g., a single memory word or 64 bits), where each pool manages a small number of variable-length counters. 
Importantly, we require an \emph{exact} counter representation to ensure that our scheme can accurately estimate the size of heavy hitters and is compatible with use cases that require exact counting, such as computing a histogram. 
This is in contrast to previous approaches that introduce errors in counter representations. For example, errors are introduced by SALSA~\cite{SALSA} when merging counters upon overflows and by PyramidSketch~\cite{PyramidSketch} when sharing the most significant bits of large counters.

The main challenge in using a pool-based solution with variable-length counters is the ability to efficiently access and update counters, adjusting their sizes whenever required.
To determine the sizes of the counters in the shared memory pool, we introduce a \emph{configuration number} that encodes it. 
We note that while the number of possible configurations is exponential in the number of counters, the number of bits required to represent a configuration is still small compared to the size of the pool itself. The overall memory required by a pool is then its memory in addition to the number of bits required for representing the configuration.
For example, if a 64-bit sized pool is shared among six counters, and the configuration size requires eight bits, then we have a total of $72/6=12$ bits per counter on average.

\smallskip
More specifically, to encode how the pool's bits are partitioned between the counters, we use the stars-and-bars combinatoric structure (see~\cite[Chapter I.3]{flajolet2009analytic}) and present efficient encoding (given counter sizes, produce a configuration number) and decoding (given a configuration number, output the counter sizes) algorithms.
%
%
Additionally, because each individual pool typically holds only a few counters, the limited number of configurations allows for the effective use of lookup tables, which speeds up processing. These tables make encoding and decoding highly efficient, and when dealing with a large array of counters, their overhead is minimal since only one copy of each table is required for the entire array.

\smallskip
We note that while our design allows moving bits around between counters within the same pool, if the input size is too large for the current memory allocation, pool failures can occur. That is, if the sum of bits required for the counters inside of a pool exceeds its memory, we cannot allocate the bits accordingly and we say that the pool has failed.
We investigate the strategies to mitigate the impact of pool failures, and the solution depends on the application. In approximate processing methods, we show how to repurpose the memory of failed pools to ensure measurement accuracy, while for exact solutions we migrate keys between different pools to improve the chances of capturing all sizes accurately.

\smallskip
When applied to approximate data structures, our dynamic resizing approach improves the classic sketches as they are composed entirely of counter arrays. For exact use cases, such as storing a histogram, we also suggest a variant of Perfect Cuckoo Filters~\cite{reviriego2021perfect} that allows compressing both the keys and counter values.
Namely, items are migrated to their other Cuckoo hash location if their current pool is about to fail. That is, unlike traditional Cuckoo hashes and filters that only move keys once their bucket is full, our solution may choose to shift items around to balance the number of bits required and avoid pool failures.
The resulting data structure requires less memory per entry and is therefore, when fixing the workload and space consumption, \emph{faster} than the alternatives as this translates to a lower load factor.  



We summarize our contributions as follows:
\begin{itemize}
    \item We suggest a novel counter-sizing algorithm that improves the space efficiency of counting alternatives. 
    \item We demonstrate how, for approximate applications such as sketches, our approach has a better space/accuracy tradeoff compared to state-of-the-art solutions such as SALSA~\cite{SALSA}, ABC~\cite{gong2017abc}, Pyramid~\cite{PyramidSketch}. The improvement is especially large for estimating the sizes of heavy hitter elements.
    \item We exemplify the applicability of our approach to exact algorithms by improving the task of computing a histogram over the data. Compared to Perfect Cuckoo Filters~\cite{reviriego2021perfect}-based tables and to off-the-shelf hash tables such as Robin Hood map~\cite{RobinMap} \texttt{std::unordered\_map}, we show how the compactness of our encoding translates to lower load factors and thus higher throughput.
    \item We release our code as open source to ensure the reproducibility of our work~\cite{counterpools}.
\end{itemize}
%

The remainder of the paper is organized as follows: Section~\ref{sec:related} surveys related work. In Section~\ref{sec:CP}, we detail our algorithm, beginning with the stars-and-bars principles and configuration number assignments in Section~\ref{sec: Configuration Number}, followed by a discussion on the storage of counters within the pool’s memory block in Section~\ref{sec:Layout}, various optimizations in Section~\ref{sec:Optimization}, and methods for handling pool failures in Section~\ref{sec:poolfailures}. Section~\ref{sec:Applications} discusses the applications of our method, including sketches in Section~\ref{sec:Sketches} and histograms in Section~\ref{sec:Histograms}. The evaluation of our algorithms and their comparison with the state of the art are presented in Section~\ref{sec:evaluation}. Finally, we conclude with~a~discussion~in~Section~\ref{sec:discussion}.

\section{Related Work}\label{sec:related}

\paragraph{\textbf{Sketch algorithms.}} Methods such as the Count-Min-Sketch~\cite{CountMinSketch}, the Count Sketch~\cite{CountSketch}, and Universal Sketch~\cite{univmon} are composed of fixed-sized shared counter arrays with multiple hash functions. Such a common structure is common across many sketch algorithms for multiple types of approximate statistics.  Specifically, the Universal Sketch~\cite{univmon,UnivMonTheory}  can estimate a large family of functions of the frequency vector, allowing the user to determine the metric of interest on the fly, or Cold Filter~\cite{ColdFilter} that optimizes accuracy and speed by filtering out cold items preventing \mbox{them from updating the sketch. }

Sketch algorithms that solve the same analytic problem (e.g., heavy hitters) are comparable by their update speed, throughput, and space/accuracy tradeoff. Namely, the more counters we fit into their counter arrays, the higher the accuracy is. Examples include Nitro~\cite{Nitro} and Randomized Counter Sharing~\cite{RandomizedCounterSharing} that trade space efficiency for update speed. As well as algorithms like
Counter Braids~\cite{CounterBraids} and Counter Tree~\cite{CounterTree} trade update speed for more accuracy where the desired approach and optimization criteria depend on the specific domain. Nitro and Randomized Counter Sharing use a form of a random sample to accelerate the runtime. In short, they perform fewer counter updates, which increases the error rate but still provides statistical guarantees. Methods that optimize space usually include a hierarchical structure that allows for dynamic sizing of counters. Examples include Counter Braids, Counter Tree, and Pyramid sketch~\cite{PyramidSketch, CounterTree,CounterBraids}, where overflowing counters are extended to a secondary data structure which allows us to size counters for the average case rather than for the worst case. However, such a hierarchical approach usually slows the computation as we perform more memory accesses and hash calculations.  
\begin{table}[t]
\resizebox{.999\columnwidth}{!}{%
\begin{tabular}{|c||l|}
\hline
\textbf{Symbol}   & \textbf{Meaning}                                                                          \\ \hline\hline
\textbf{$n$}      & The number of bits per pool.                                                  \\ \hline
\textbf{$k$} & The number of counters per pool.                                                     \\ \hline
\multirow{2}{*}{\textbf{$SnB(n,k)$}} & The number of options to place $n$ identical \\   & balls into $k$ distinguishable bins.                                                     \\ \hline

\textbf{$x_1,\ldots,x_k$} & An encoded configuration.                                                     \\ \hline

\multirow{2}{*}{\textbf{$\xi$}} & The number of stars-and-bars configuration \\ & whose first entry is smaller than $x_0$.                                                     \\ \hline

\multirow{1}{*}{$T[a,b,c]$} & A lookup table value of $\sum_{j=0}^{c-1} SnB(a-j,b-1)$.                                                     \\ \hline

\multirow{1}{*}{$C$} & A configuration number.                                                     \\ \hline

\multirow{1}{*}{$\rho$} & The first decoded value given $C,n,k$.                                                     \\ \hline

\multirow{1}{*}{$\psi$} & The other decoded values given $C,n,k$.                                                     \\ \hline

\multirow{1}{*}{$s$} & The starting size of each counter.                                                    \\ \hline
\multirow{1}{*}{$i$} & The number of bits added when a counter overflows.                                                    \\ \hline
\multirow{1}{*}{$w$} & The weight added to a counter in an Increment.                                                    \\ \hline

\multirow{2}{*}{$L$} & maps each $C\in SnB(64,4)$ into an array of $k=4$ \emph{offsets}\\ & of where each counter starts in the pool's memory block.                                                    \\ \hline
\end{tabular}
}
\vspace{0mm}
\caption{The notations used in the paper.}\label{tbl:notations}
\vspace{-4mm}
\end{table}

Other algorithms suggest ways for overflowing counters to grow at the expense of other counters; for example, in ABC~\cite{gong2017abc}, an overflowing counter `steals' a bit from its subsequent counter, whereas in SALSA~\cite{SALSA}, it merges with a neighboring counter forming a double-sized counter. Our work follows this flavor but is more ambitious than ABC and SALSA. Namely, we distribute the bits of a pool according to demand rather than stealing bits from a neighboring counter (which then can steal more bits from its neighbor). Such an on-demand approach yields a lower chance of overflow and faster runtime. SALSA's approach of merging overflowing counters only applies to approximate shared counter solutions, whereas our approach applies to accurate streaming algorithms. Even when considering the approximate algorithms, merging counters increases the error of the largest heavy hitter flows, which is usually undesirable.
In contrast,  up-sizing counters without a merge does not increase the error. However, unlike SALSA, our approach yields overflows that must be handled (e.g., by employing secondary structures). Here, we consider multiple options and explain their impact. The secret sauce is in balancing these \mbox{approaches to yield an attractive tradeoff. }

Other approaches use multiple forms of sampling to reduce the counter length by only incrementing counters with a certain probability which is either fixed~\cite{AEE} or depends on the specific counter value~\cite{CEDAR,ANLSUpscaling,ApproximateCounting}. Such approaches save space by counting to larger numbers with a small counter size, but their use of sampling does not apply to all measurement tasks. Moreso, they complement our approach because even sampled counters need to be stored in memory and are expected to have variations in values.

\paragraph{\textbf{Histograms.}} To the best of our knowledge, the task of computing (an exact) histogram succinctly is novel to our paper, and current approaches typically use a simple hash table (such as the Robin Hash table~\cite{RobinMap} for that purpose. Also, to have an additional strong baseline, we extend the Perfect Cuckoo Filter algorithm~\cite{reviriego2021perfect} to support holding values.

\section{Counter Pools}\label{sec:CP}
This section presents the counter pools framework for variable counter sizes. The following subsections provide explanations of the architecture. Section~\ref{sec: Configuration Number} defines the encoding and decoding, while Section~\ref{sec:Layout} shows the counter layout. Then, we discuss optimizations in Section~\ref{sec:Optimization} and provide an illustrated example of how everything is connected. We summarize the notations used in this paper in~\Cref{tbl:notations}.

\subsection{Pool configuration number}
\label{sec: Configuration Number}

This subsection explains how we translate a small subset of counter sizes into a configuration number and vice versa.
Consider representing the counter sizes in a pool of $n$ bits shared among $k$ counters. Let us assume that counters start with $0$ bits (i.e., can represent only the value $0$) and grow by a bit at a time whenever needed. The pool is said to be \emph{functioning} as long as the sum of counter sizes does not exceed $n$.

The number of possible size configurations, in this case, is the stars-and-bars~\cite[Chapter I.3]{flajolet2009analytic} value $SnB(n,k+1) = {n+k \choose k}$, which is the number of options of placing $n$ balls into $k+1$ bins; here, we use the value $k+1$ instead of $k$ since the sum of sizes does not need to reach $n$, so we have ``an extra bin'' that accounts for the number of unallocated bits.
The main ingredient of efficiently implementing the encoding and decoding of stars-and-bars is converting a sequence of $k$ sizes that sum up to $n$ into a configuration number in $\set{0,\ldots,SnB(n,k+1)-1}$ and vice versa. 

Given a fixed integer $k$ and a non-negative integer sum $n$, we establish a bijection between the set of all $k$-partitions of $n$ and a subset of non-negative integers. A $k$-partition of $n$ is defined as a list of $k$ non-negative integers that sum to $n$. The bijection is constructed using combinatorial principles, specifically leveraging stars and bars as described below.
Specifically, the encoding process transforms a given $ k$ partition into a unique integer by interpreting the partition as a distribution of $n$ indistinguishable items into $k$ distinguishable bins. This encoding uses a sequence of $n$ items and $k-1$ dividers. The position of the partition in a lexicographically ordered set of all partitions corresponds to the bijection number, calculated via the sum of combination \mbox{counts preceding the given partition.}

The inverse decode operation reconstructs the original $k$-partition from the bijection number by performing a reverse lexicographic unranking. This decoding involves iteratively determining the next integer. The process continues until the entire $k$-partition is recovered.
This bijection allows for a compact integer representation of $k$-partitions of $n$, facilitating efficient storage and computation. Moreover, the one-to-one correspondence ensures that each partition is uniquely represented by a single integer and vice versa, allowing for direct access and manipulation \mbox{within combinatorial algorithms.}

For the encoding process, which is given in Algorithm~\ref{alg:Encode}, we first calculate $\xi$ (Line~\ref{line:xi}), which is the number of stars-and-bars configuration whose first entry is smaller than $x_0$. We then add to it the value returned by the recursive call $\text{Encode}([x_1,\ldots,x_{k-1}])$. The correctness of the algorithm is a simple inductive argument that observes that all the indices of sequences that start with $x$ are between $\sum_{j=0}^{x_0-1}SnB(-j+\sum_{\ell=0}^{k-1}x_\ell , k-1)$ and $\sum_{j=0}^{x_0}SnB(-j+\sum_{\ell=0}^{k-1}x_\ell , k-1)-1$.

\begin{algorithm}[tb]
   \caption{Encode$([x_0,\ldots,x_{k-1}], n)$}
   \label{alg:Encode}
\begin{algorithmic}[1]
    \If{$k=1$}
        \State\textbf{Return $0$}\label{line:return0}
    \EndIf
    \State $\xi = \sum_{j=0}^{x_0-1}SnB(n-j, k-1)$\label{line:xi}
    \State\textbf{Return}
    $\text{Encode}([x_1,\ldots,x_{k-1}], n-x_0) + \xi$

\end{algorithmic}
\end{algorithm}

\begin{algorithm}[tb]
   \caption{Decode$(C, n, k)$ }
   \label{alg:Decode}
\begin{algorithmic}[1]
    \If{$k=1$}
        \State\textbf{Return $[n]$}
    \EndIf
    \State $\rho = 0$\label{line:rho0}
    \If {$C>0$}
        \State $\rho = \max\set{x \mid \sum_{j=0}^x SnB(n - j, k-1) \le  C}$\label{line:rho}
    \EndIf
    \State $\psi = \text{Decode}(C - \sum_{j=0}^{\rho-1}SnB(n - j, k-1), n-\rho, k-1)$
    \State\textbf{Return}
    $\text{Concat}([\rho], \psi)$

\end{algorithmic}
\end{algorithm}

\begin{algorithm}[tb]
   \caption{Encode$([x_0,\ldots,x_{k-1}], n)$ with lookup}
   \label{alg:EncodeWlookup}
\begin{algorithmic}[1]
    \If{$k=1$}
        \State\textbf{Return $0$}
    \EndIf
    \State $\xi = T[n, k-1, x_0]$
    \State\textbf{Return}
    $\text{Encode}([x_1,\ldots,x_{k-1}], n-x_0) + \xi$
\end{algorithmic}
\end{algorithm}

\begin{algorithm}[tb]
   \caption{Decode$(C, n, k)$ with lookup}
   \label{alg:DecodeWlookup}
\begin{algorithmic}[1]
    \If{$k=1$}
        \State\textbf{Return $[n]$}
    \EndIf
    \State $\rho = 0$
    \While{$T[n, k-1, \rho+1] \le C$}
        \State $\rho = \rho + 1$
    \EndWhile
    \State $\psi = \text{Decode}(C - T[n, k-1, \rho], n-\rho, k-1)$
    \State\textbf{Return}
    $\text{Concat}([\rho], \psi)$

\end{algorithmic}
\end{algorithm}

\renewcommand{\arraystretch}{1.6}
\begin{table*}[t]
\centering
\begin{tabular}{|l|l|l|l|}
\hline
\textbf{Input} & \textbf{$\xi$ Computation} & \textbf{$\xi$ Value} & \textbf{Recursive Call} \\ \hline
[26, 20, 8, 0, 10], 64 & $\sum_{j=0}^{26-1}SnB(64-j, 5-1)$ & 702455 & $\text{Encode}([20,8,0,10], 64-26)$ \\ \hline
[20, 8, 0, 10], 38     & $\sum_{j=0}^{20-1}SnB(38-j, 4-1)$ & 9330   & $\text{Encode}([8,0,10], 38-20)$ \\ \hline
[8, 0, 10], 18         & $\sum_{j=0}^{8-1}SnB(18-j, 3-1)$  & 124    & $\text{Encode}([0,10], 18-8)$    \\ \hline
[0, 10], 10            & $\sum_{j=0}^{-1}SnB(10-j, 2-1)$   & 0      & $\text{Encode}([10], 10)$        \\ \hline
[10], 10               & Return 0                          & 0     & -                                \\ \hline
\end{tabular}
\vspace{1mm}
\caption{Encoding steps and calculations. The return value for the 5-partition $[26, 20, 8, 0, 10]$ is the sum of all $\xi$ values: $C=702455+9330+124+0+0=711909$.}
\vspace{-2mm}
\label{tab:encoding_steps}
\end{table*}

\renewcommand{\arraystretch}{1.6}
\begin{table*}[t]
\centering
\begin{tabular}{|l|l|l|l|}
\hline
\textbf{Input}  & \textbf{$\rho$ Computation} & \textbf{$\rho$ value} & \textbf{Recursive Call} \\ \hline
$711909, 64, 5$ & $\max\set{x \mid \sum_{j=0}^x SnB(64 - j, 5-1) \le  711909}$ & $\rho = 26$ & $\text{Decode}(711909 - \sum_{j=0}^{26-1} SnB(64 - j, 5-1), 64-26, 5-1)$ \\ \hline
$9454,   38, 4$ & $\max\set{x \mid \sum_{j=0}^x SnB(38 - j, 4-1) \le  9454  }$ & $\rho = 20$ & $\text{Decode}(9454   - \sum_{j=0}^{20-1} SnB(38 - j, 4-1), 38-20, 4-1)$ \\ \hline
$124,    18, 3$ & $\max\set{x \mid \sum_{j=0}^x SnB(18 - j, 3-1) \le  124   }$ & $\rho = 8 $ & $\text{Decode}(124    - \sum_{j=0}^{8 -1} SnB(18 - j, 3-1), 18-8 , 3-1)$ \\ \hline
$0,      10, 2$ & $ \rho = 0$  (since $C=0$)& $\rho = 0 $& $\text{Decode}(0      - \sum_{j=0}^{0 -1} SnB(10 - j, 2-1), 10-0 , 2-1)$ \\ \hline
$0,      10, 1$ & Return [10]  & $\rho = 10$ &  - \\ \hline
\end{tabular}
\vspace{3mm}
\caption{Recursive calls in the decode function. The return value for $C=711909, n=64, k=5$ is the sequence of the $\rho$ values: $[26, 20, 8, 0, 10]$.}
\vspace{-2mm}
\label{tab:decode_calls}
\end{table*}

   \begin{figure*}[t]
        \subfloat[Basic Layout]{\label{fig:layout1}
        \includegraphics[width =0.49493405\linewidth]{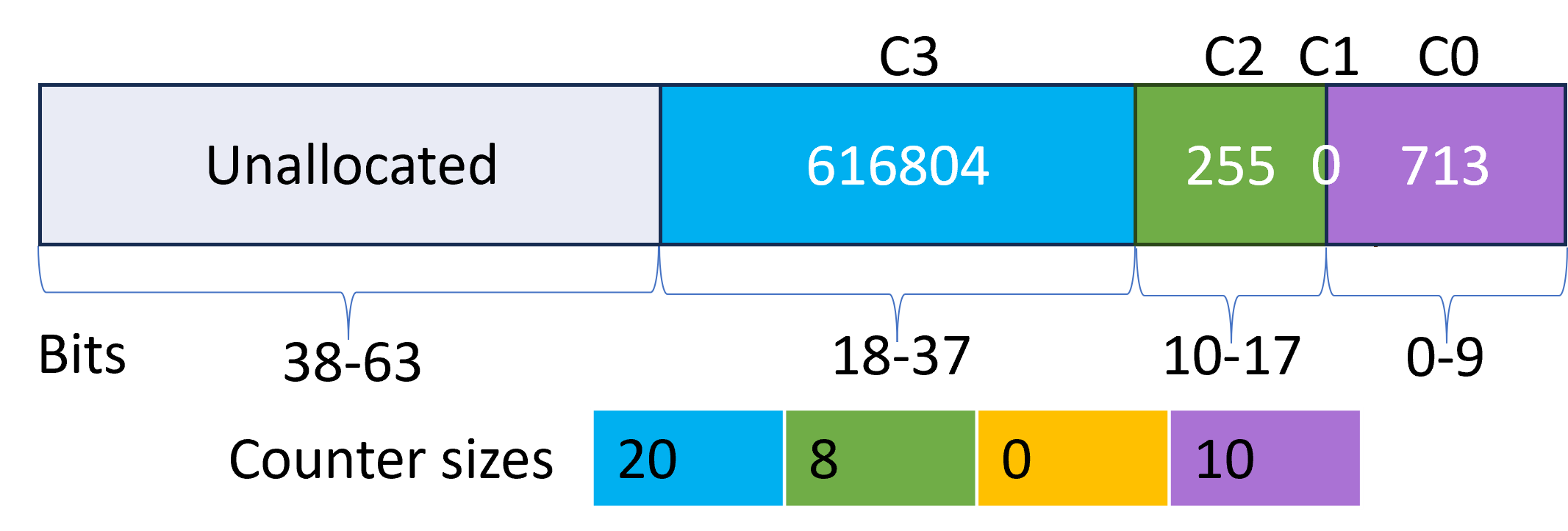}}
        \subfloat[After Incrementing C2]{\label{fig:layout2}
        \includegraphics[width =0.49493405\linewidth]{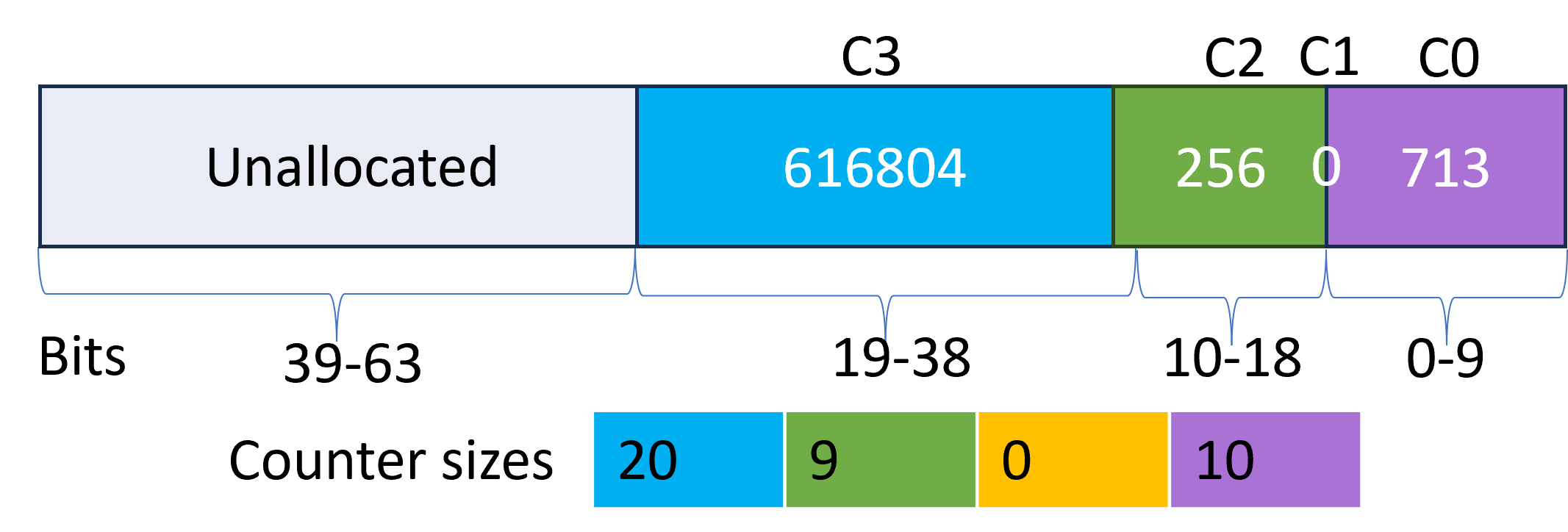}}
        \centering
        
        \caption{Basic Layout of the counters and an increment operation.\label{fig:counter_pools_conf}}
      \end{figure*}      

The decoding process, given by Algorithm~\ref{alg:Decode} works analogously -- it first infers the value of the first entry $\rho$ (Line~\ref{line:rho}) using the above observation and then recursively calculates the remaining entries based on the residual configuration number. To allow an efficient calculation, one can use a lookup table that stores $T[a,b,c] = \sum_{j=0}^{c-1} SnB(a-j,b-1)$ for all $a\in\set{1,\ldots,n},b\in\set{1,\ldots,k},c\in\set{1,\ldots,a}$. This lookup table requires $O(n^2 \cdot k)$ space and allows $O(k)$ time encoding and $O(n+k)$ time decoding. Here, $T$ is calculated offline, and its initialization does not impact the encoding and decoding time.
The pseudo codes that implement this optimization are given by Algorithm~\ref{alg:EncodeWlookup} and Algorithm~\ref{alg:DecodeWlookup}. 

In Algorithm~\ref{alg:DecodeWlookup}, we also propose a specific way to compute $\rho$ (Line 3 in~\ref{alg:Decode})\footnote{There are different ways to calculate $\rho$ such as using a binary search or a predecessor search data structure. The complexities of such approaches are between $O(k\log\log n)$ and $O(k\log n)$, and thus most efficient way would depend on the values of $k$ and $n$.} that work well for different values of $k$ and $n$ we are interested in. The decoding time is then established by observing that the while loop (Line 4) can be executed at most $n$ times in all recursive calls combined. This is because at each loop iteration, $\rho$ is incremented by 1, and the second argument \mbox{in the recursive call is $n-\rho$.}

For example, consider the $5$-partition $(26, 20, 8, 0, 10)$ illustrated in~\Cref{fig:layout1} (for $n=64$).
The encoding process is shown in \Cref{tab:encoding_steps}; each line represents an invocation of the function and calculates the value of $\xi$ before calling the next line recursively. Similarly, the decoding process, given in~\Cref{tab:decode_calls}, finds the value of $\rho$ before invoking the next line recursively. For example, in the first row, we have $\sum_{j=0}^{26} SnB(64 - j, 5-1) = 702455\le C$ and $\sum_{j=0}^{27} SnB(64 - j, 5-1) = 713115 > C$ and thus $\rho=26$.

        \begin{figure*}[t!]
        \subfloat[Basic Layout]{\label{fig:layout3}
        \includegraphics[width =0.49493405\linewidth]{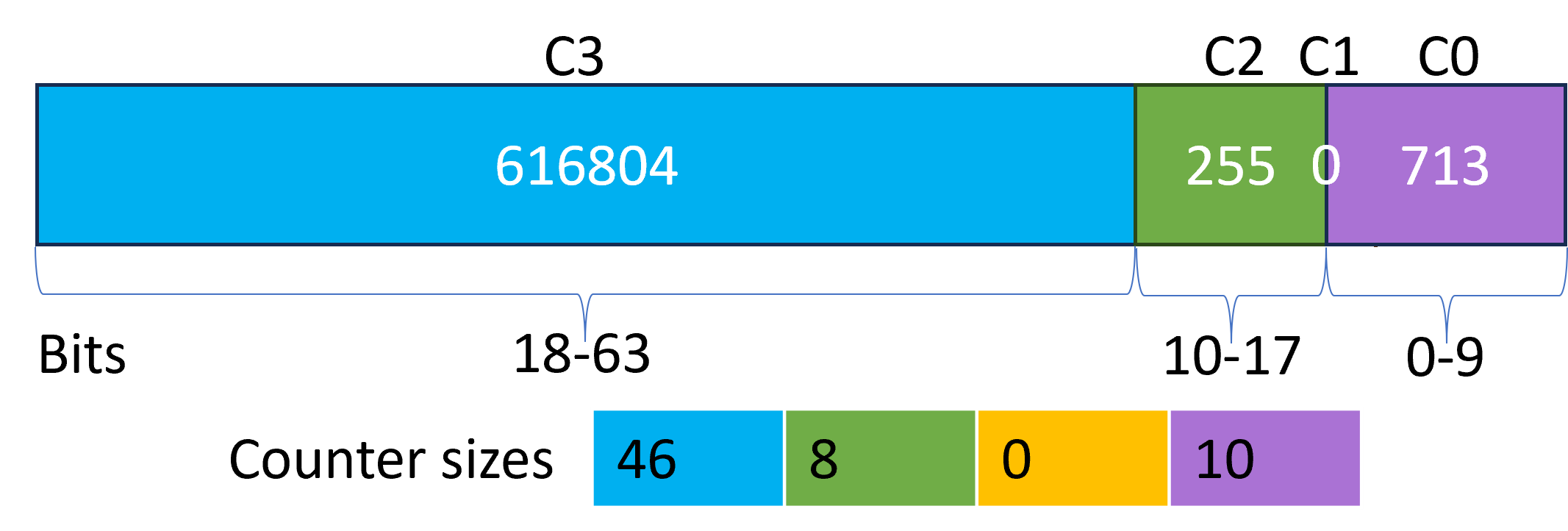}}
        \subfloat[After Incrementing C2]{\label{fig:layout4}
        \includegraphics[width =0.49493405\linewidth]{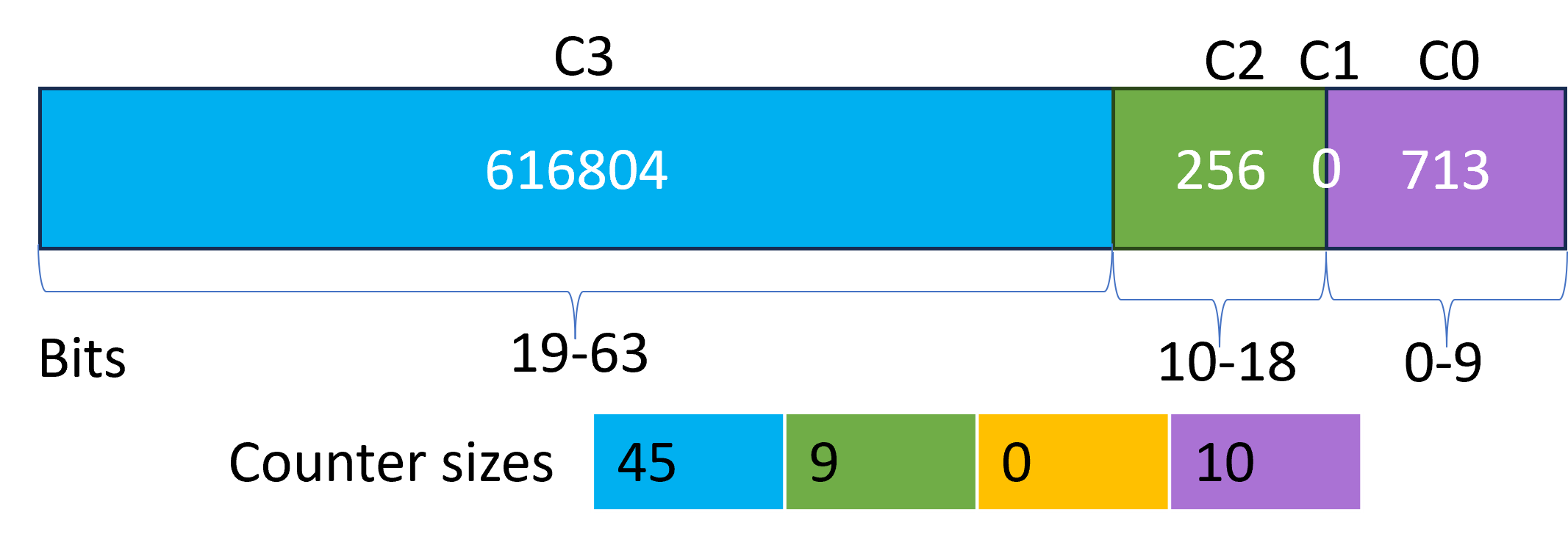}}
        \centering
        \vspace*{1mm}
        
        \caption{Optimization: placing unallocated bits at the leftmost counter.\label{fig:counter_pools_conf2}}
        \vspace*{1mm}
        \end{figure*}  

\subsection{Counters layout}
\label{sec:Layout}

In our algorithms, each pool is represented consecutively in memory using a consecutive chunk of $n$ bits in addition to a configuration number $C\in SnB(n,k+1)$. The counters are left-aligned, e.g.,  if the counter sizes are $(20,8,0,10)$, as illustrated in Figure~\ref{fig:layout1}, and (e.g., the pool is represented using a single \texttt{uint64\_t} variable), then the ten least significant bits belong to the first counter, bits $10-17$ belong to the third one (i.e., the second counter does not have any bits), the next $20$ bits belong to the fourth, \mbox{and the rest are unallocated. }

Suppose, for the sake of the example, that we now need to increment counter C2 to have the value $256$. 
The first step is to look at the current counter sizes (which are encoded with the configuration number) and assess whether the new value can fit.
Since this cannot be represented using the currently allocated eight bits, we need to allocate it with an additional bit. As illustrated in Figure~\ref{fig:layout2}, this may require shifting some counters (C3, in the example) to keep the counters left-aligned, but is done in a constant number of operations (that is independent of the number of counters in the pool). If there is no available unallocated bit, we say that the pool has \emph{failed}. Section~\ref{sec:poolfailures} explains how we handle such pool failures.




%

\begin{algorithm}[tb]
   \caption{AccessCounter$(\mathit{pool}, i)$ with lookup}
   \label{alg:Access}
\begin{algorithmic}[1]
    \State $\mathit{offset} = L[\mathit{pool}.C, i]$ 
    \State $\mathit{next\_offset} = L[\mathit{pool}.C, i+1]$ 
    \State $\mathit{size} = \mathit{next\_offset}-\mathit{offset}$ 
    \State\textbf{Return} $(\mathit{pool.M}  \gg \mathit{offset})\ \&\ (2^{size}-1)$
\end{algorithmic}
\end{algorithm}

\begin{algorithm}[tb]
   \caption{IncrementCounter$(\mathit{pool}, i, w)$ with lookup}
   \label{alg:IncCounter}
\begin{algorithmic}[1]
    \State $\mathit{offset} = L[\mathit{pool}.C, i]$ \label{line:calcOffset}
    \State $\mathit{next\_offset} = L[\mathit{pool}.C, i+1]$ \label{line:calcNextOffset}
    \State $\mathit{size} = \mathit{next\_offset}-\mathit{offset}$  \label{line:csize}
    \State $\mathit{v} = (\mathit{pool}.M  \gg \mathit{offset})\ \&\ (2^{\mathit{size}}-1)$ \label{line:val}
    \If{$2^{\mathit{size}-1} \le v + w < 2^{\mathit{size}}$}\label{line:canfit}
        \State $\mathit{mask} =\ \ \sim((2^{size}-1) \ll \mathit{offset})$ \label{line:mask}
        \State\textbf{$\mathit{pool}.M = ((v + w) \ll \mathit{offset})\ | \ (\mathit{pool}.M\ \&\ \mathit{mask})$}\label{line:updateVal}
        \State\textbf{Return}
    \EndIf
    \State $\mathit{new\_size} = \ceil{\log_2{(v + w + 1)}}$\label{line:newsize}
    \State $\mathit{new\_bits} = \mathit{new\_size} - \mathit{size}$\label{line:newbits1}
    \State $\mathit{lc\_offset} = L[\mathit{pool}.C, k-1]$  \Comment{Last counter's offset} \label{line:lastoffset}
    \State $\mathit{lc\_size} = n-\mathit{lc\_offset}$ \label{line:lastsize}
    \State $\mathit{lc\_required\_size} = \ceil{\log_2({1+(\mathit{pool}.M \gg \mathit{lc\_offset}}))}$\label{line:lastrequiredsize}
    \State $\mathit{free\_bits} = \mathit{lc\_size}-\mathit{lc\_required\_size}$\label{line:freebits}
    \If{$\mathit{new\_bits} > \mathit{free\_bits}$}\label{line:poolFailure}
        \State\textbf{Return} Pool failure \Comment{See Section~\ref{sec:poolfailures}}
    \EndIf
    \State \textbf{$\mathit{low} = \mathit{pool}.M\ \&\ (2^{\mathit{offset}}-1)$}\label{line:lowbits}
        \State \textbf{$\mathit{new} = (v+w)\ \ll\ ({\mathit{offset}})$}\label{line:newbits}
        \State \textbf{\small{$\mathit{high} = (\mathit{pool}.M\gg\mathit{next\_offset})\ll(\mathit{next\_offset} + \mathit{new\_bits})$}}\label{line:highbits}
        \State\textbf{$\mathit{pool}.M = \mathit{high} \ | \ \mathit{new}\ |\ \mathit{low}$}\label{line:newM}
        \For{$j{\in}\set{0,\ldots,k-2}$}\label{line:oldCoutnerSizes}
            \If{$j\neq i$}
                \State $\mathit{size}_j = L[\mathit{pool}.C, j+1] - L[\mathit{pool}.C, j]$
            \EndIf
        \EndFor
        \State $\mathit{size}_i = \mathit{new\_size}$\label{line:sizei}
        \State $ \mathit{size}_{k-1} = \mathit{lc\_size} -\mathit{new\_bits}$\label{line:sizek-1}
        \State $\mathit{pool}.C = \mathit{Encode}(\set{size_i}_{i=0}^{k-1}, n)$\Comment{Algorithm~\ref{alg:EncodeWlookup}}\label{line:newC}
        \State\textbf{Return}
\end{algorithmic}
\end{algorithm}

\subsection{Optimizations}
\label{sec:Optimization}

\paragraph{Reducing the encoding overheads}
\label{sec:ReducedEncoding}

Heretofore, we described the unallocated bits as sitting in a separate memory space (as illustrated in Figure~\ref{fig:counter_pools_conf}). As a result, the number of possible configurations for a pool is given by $SnB(n,k+1)$.
For example, with $k=4$ counters sharing $n=64$ bits, we have $SnB(64,5)=814385$, which means that one would require $\ceil{\log_2 SnB(64,5)}=20$ bits to represent the current configuration.

Instead, we change the layout to have all the unallocated bits inside of the leftmost counter. Consider the previous example for the counter values. As illustrated in Figure~\ref{fig:counter_pools_conf2}, C3 holds with all the pool's bits except those allocated for C0-C2 (Figure~\ref{fig:layout3}). When a counter requires additional bits, as in the case of incrementing C2, a bit from C3 is taken if it has one to spare (and otherwise, the pool fails. A simple observation is that the optimization does not increase the probability of pool failures, as the same access pattern would cause a pool failure in both layouts or in none of them.
The benefit from the change is that now we only have $SnB(64,4)=47905$ configurations, and thus we can fit the configuration number into a $16$-bit~variable.

A second approach for reducing the number of configurations (and thus the encoding overheads) is to allow counters to start from more than $0$ bits or to allocate counters with more than one additional bit when the current allocation is not enough.
Namely, we define a Counter Pools configuration as \mbox{a 4-tuple $(n,k,s,i)$, such that:}
\begin{itemize}
    \item $n$ is the number of bits of the pool.
    \item $k$ is the number of counters.
    \item $s$ is the starting size of each counter.
    \item $i$ is the number of bits we allocate whenever increasing the width of a counter.
\end{itemize}
For example, the above configuration that started with $s=0$ bits and allocating $i=1$ bit at a time would be denoted as $(64,4,0,1)$.

While this allows less flexibility (and thus potentially increases the probability of pool failures), it reduces the number of required configurations to $SnB(\floor{\frac{n-k\cdot s}{i}},k)$.
For example, with $k=4$ counters sharing $n=64$ bits, if we use $s=12, i=2$, the number of configurations reduces to $SnB(8,4)=165$, which can be represented using an $8$-bit variable.
As additional examples, it allows for fitting $k=5$ (e.g., with $s=8,i=4$) or $k=6$ (e.g., with $s=7,i=4$) counters within $n=64$ bits using $8$-bit configurations.

\paragraph{Accelerating the counter accesses}
As described above, one can access a given counter by first decoding the counter sizes in $O(n+k)$ time and calculating the offset of the counter.
However, in many configurations, we can significantly accelerate the accesses by using an additional lookup table.
For example, consider the configuration $(64,4 0,1)$. We can create a lookup table $L$ that maps each $C\in SnB(64,4)$ into an array of $k=4$ \emph{offsets} that indicate where each counter starts in the pool's memory block. For example, in Figure~\ref{fig:layout3}, the counter sizes for $(C0,C1,C2,C3)$ were $(10,0,8,46)$ and thus the offsets are $(0,10,10,18)$.
The lookup table, in this example, has $SnB(64,4)$ entries that can be encoded using 32-bit each (as each offset requires at most six bits) for a total of 187KB, which are not tied to a specific pool and can be shared between multiple pools (e.g., in a sketch, or even multiple sketches).

Given a configuration number $C$, we can then access and update a counter $c\in\set{0,1,2,3}$ in constant time: we first use the lookup table $L$ to calculate the offsets of $c$ and $c+1$\footnote{We denote counter 4's offset as $64$ to allow the same logic to apply to the last counter.} which gives the specific bits of the counter.
For example, if we want to read C2 in Figure~\ref{fig:layout3}, we use the lookup table to get its offset ($10$) and the offset of C3 ($18$), which together indicates that C2 is stored in bits $10-17$.
This means that accessing a counter is done in constant time.

Importantly, when a counter requires additional bits, we only need to modify its width and that of the leftmost counter, thus also allowing changing the layout in constant time that is independent of $k$. As mentioned earlier and illustrated in Figure~\ref{fig:counter_pools_conf2}, this is achieved by allocating additional $i$ bits to the counter and shifting the counters to the left of $c$ accordingly.

More generally, there will be $SnB(\floor{\frac{n-k\cdot s}{i}},k)$ entries in such a table, which is exponential in $k$, and therefore this optimization is suitable for small values of $n,k$.

\paragraph{Putting it all together}
We now describe how to increase the value of the $i$'th counter in $\mathit{pool}$ by some $w\in\mathbb Z$, as described in~\Cref{alg:IncCounter}.
The first step (\Cref{line:calcOffset}) is to lookup the bit offset, given the configuration number $\mathit{pool.C}$. 
Next, we lookup (\Cref{line:calcNextOffset}) the offset of the $i+1$'th counter (we implicitly define $L[C, k+1]=n$ for any $C$ to allow this to work for the last counter). The subtraction of the former from the latter gives the counter size (\Cref{line:csize}).
We then calculate the current value of the counter by shifting the pool's bits and zeroing out the bits of other counters (\Cref{line:val}).
Next, we check if the current number of bits assigned to this counter is sufficient and necessary to represent its new value ($v+w$) (\Cref{line:canfit}).
If so, we calculate a mask that zeros out the bits of all other counters (\Cref{line:mask}) and use it to update the value of the counter (\Cref{line:updateVal}).
Otherwise, we have to resize the counter, and we start by calculating the required number of bits (\Cref{line:newsize}) and how many bits we need to allocate (or deallocate), as shown in~\Cref{line:newbits}. The next step is to calculate the offset (\Cref{line:lastoffset}), size (\Cref{line:lastsize}), how many bits it requires (\Cref{line:lastrequiredsize}), and then how many free bits the pool has (\Cref{line:freebits}).
If there are not enough free bits to accommodate the counter update, we fail the pool in \Cref{sec:poolfailures}.
Otherwise, we need to update the number of bits allocated to the last counter and allocate or deallocate additional ones for the $i$'th counter. This is achieved by calculating the bits of counters $0,\ldots,i-1$ (\Cref{line:lowbits}), the bits of the new counter (\Cref{line:newbits}), and the bits of counters $i+1,\ldots,k-1$ after shifting them appropriately (\Cref{line:highbits}). The resulting memory chunk is then the bitwise-or of the three parts (\Cref{line:newM}). The last step is to encode the new sizes by modifying the configuration number; this is achieved by calculating the sizes of all counters other than $i$ and $k-1$ (\Cref{{line:oldCoutnerSizes}}), the sizes of these two (\Cref{line:sizei}-\Cref{line:sizek-1}), and calling the Encode function (\Cref{line:newC}). Interestingly, the algorithm seamlessly works also when $w$ is negative, and it allows deallocating bits from $C_i$ and giving them back to $C_{k-1}$.

\paragraph{Example}
Let us explain the algorithm's steps with the example of~\Cref{fig:counter_pools_conf2}, where $i=2$ and $w=1$.
In this example, the configuration number is $46699$, corresponding to counter sizes of $46,8,0,10$, and thus the offset of $C_i$ is $10$.
As the current value of the counter ($v=255$) plus the update ($w=1$) cannot fit within the current number of bits ($8$), we require memory allocation.
The new size is then $9$ bits, meaning we require one additional bit. Since the last counter has a value of $616804$, which can fit into $20$ bits, it has more than enough to give away one bit to C2. It then calculates the new memory value (($616804 \ll 19) | (256 \ll 10) | 713 = $0x4b4b2402c9) and counter sizes (45,9,0,10) which translates to updating the configuration number of $46509$.\footnote{Here, $\ll$ represent a bitwise left-shift. For example, $616804 \ll 19=616804\cdot 2^{19}=0x4b4b200000$ and $256 \ll 10=256\cdot 2^{10} = 0x40000$.}

\subsection{Handling Pool Failures}\label{sec:poolfailures}
\vspace{1mm}


As mentioned, although our architecture facilitates the redistribution of bits among counters within the same pool, excessive input sizes may surpass the allocated memory, resulting in pool failures. 
The method of handling pool failures depends on the target application, as we describe next.

For approximate algorithms such as sketches, we can allow certain pools to fail at the cost of a controllable increase in the sketch's error. We explore different approaches such as offloading the overflowing pools to a secondary data structure or modifying the pool to contain fewer counters. We compare the different alternatives quantitatively (\Cref{app:cp_setup}) to make a data-driven design choice. 

When required for accurate counter-representation, we must not have any memory shared between counters. Instead, we explore the notion of migrating elements to different counters to ensure that \emph{all} counters have a sufficient number of bits allocated. For example, for the histogram use case, we store items in a Cuckoo hash table and allocate a pool for each row. Whenever an item arrives and incrementing its counter requires an additional bit that cannot be allocated by the pool, one of the items is kicked to its other bucket, potentially triggering a chain of eviction until all counters can fit within the memory limits of their corresponding pools.


        \begin{figure*}[t!]
        \subfloat[NYC]{
        \includegraphics[width =0.490493405\columnwidth]{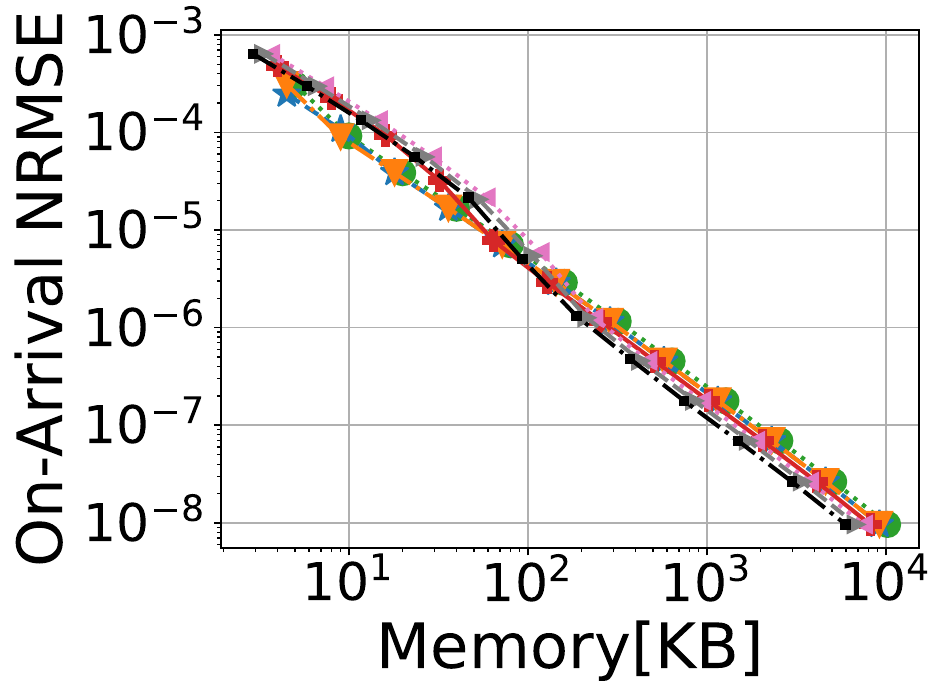}}
        \subfloat[ZIPF 0.6]
        {\includegraphics[width =0.490493405\columnwidth]
        {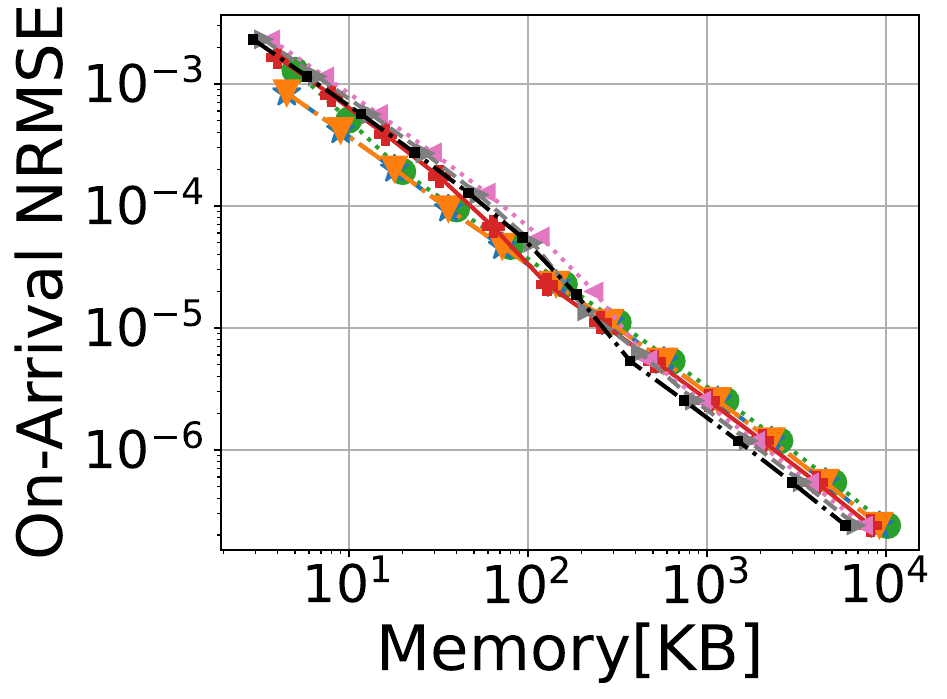}}\vspace*{3mm} 
        \centering
        \subfloat[ZIPF 1.0]
        {\includegraphics[width =0.490493405\columnwidth]{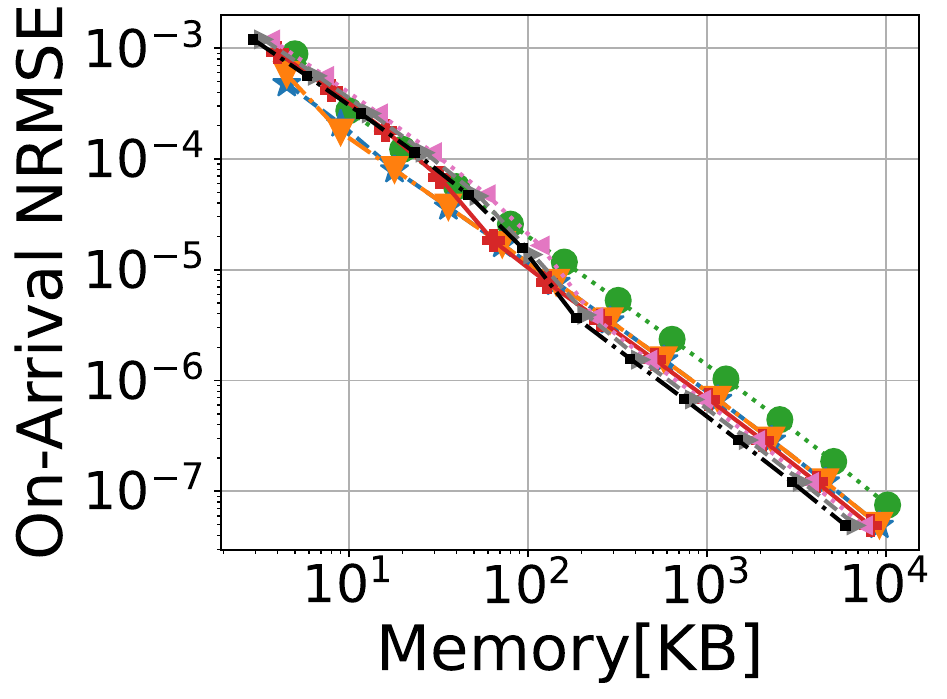}}
        \subfloat[ZIPF 1.4]
        {\includegraphics[width =0.490493405\columnwidth]
        {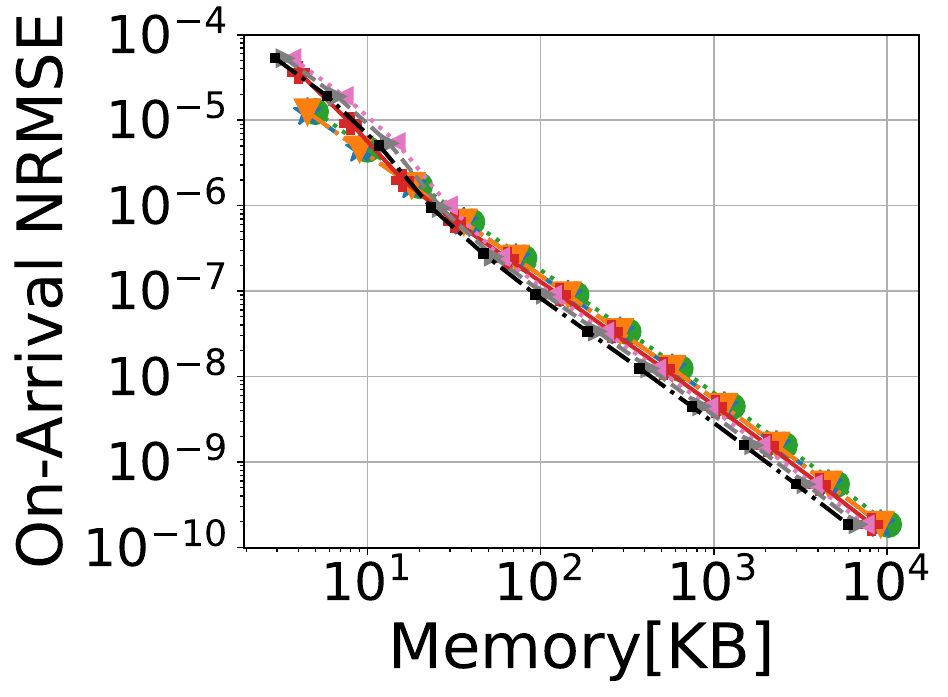}}
    
        {\includegraphics[width =2\columnwidth]
        {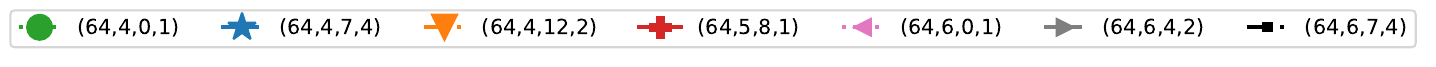}}
    
         \subfloat[NYC (zoomed)]{
        \includegraphics[width =0.490493405\columnwidth]{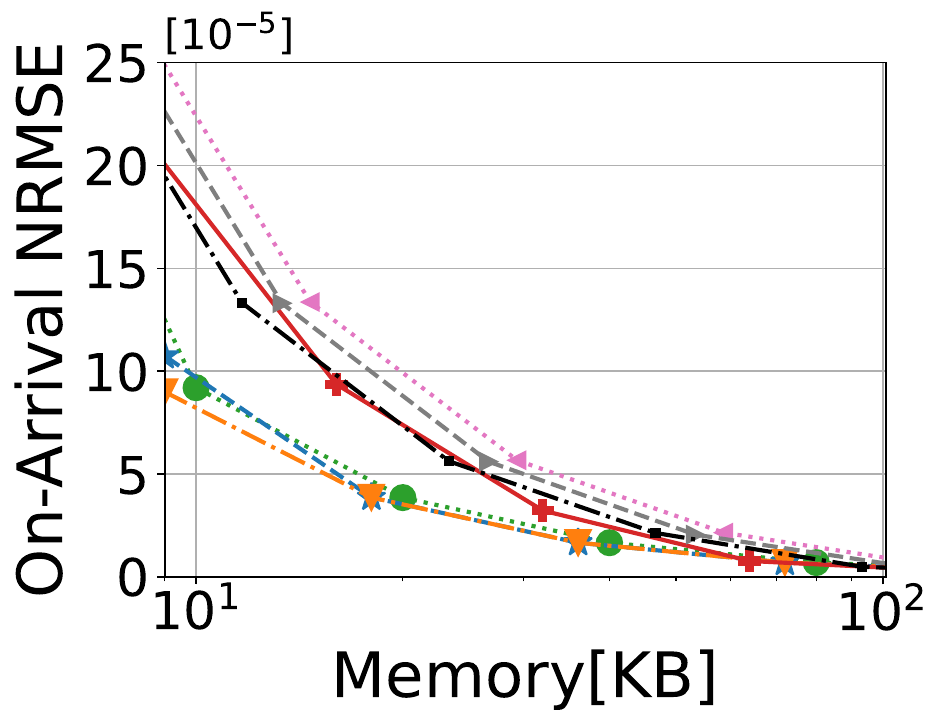}}
        \subfloat[ZIPF 0.6 (zoomed)]
        {\includegraphics[width =0.490493405\columnwidth]
        {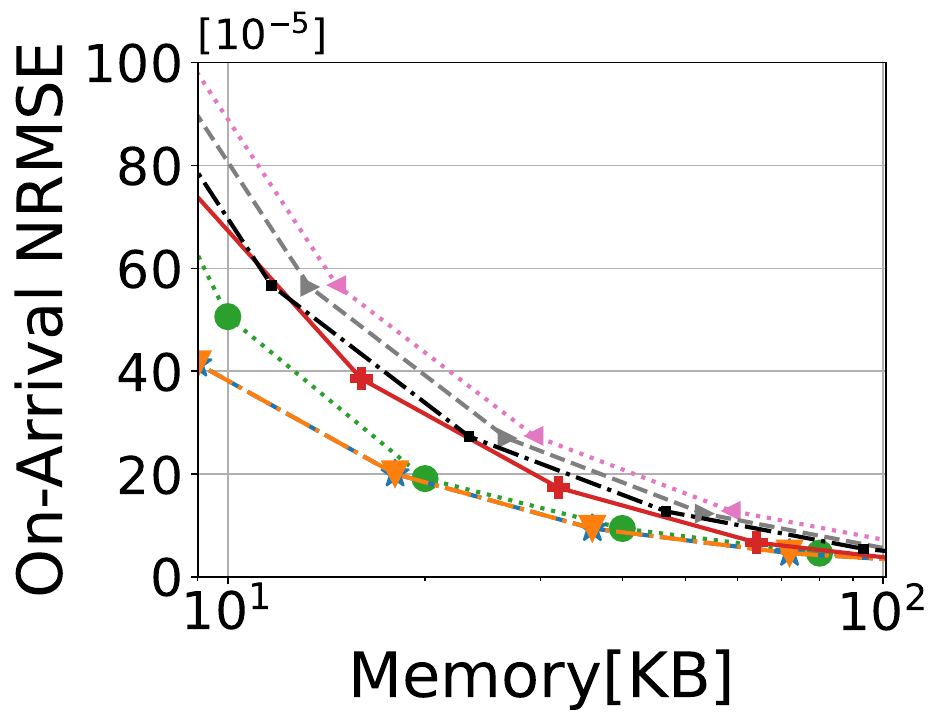}}
        \centering
        \subfloat[ZIPF 1.0 (zoomed)]
        {\includegraphics[width =0.490493405\columnwidth]{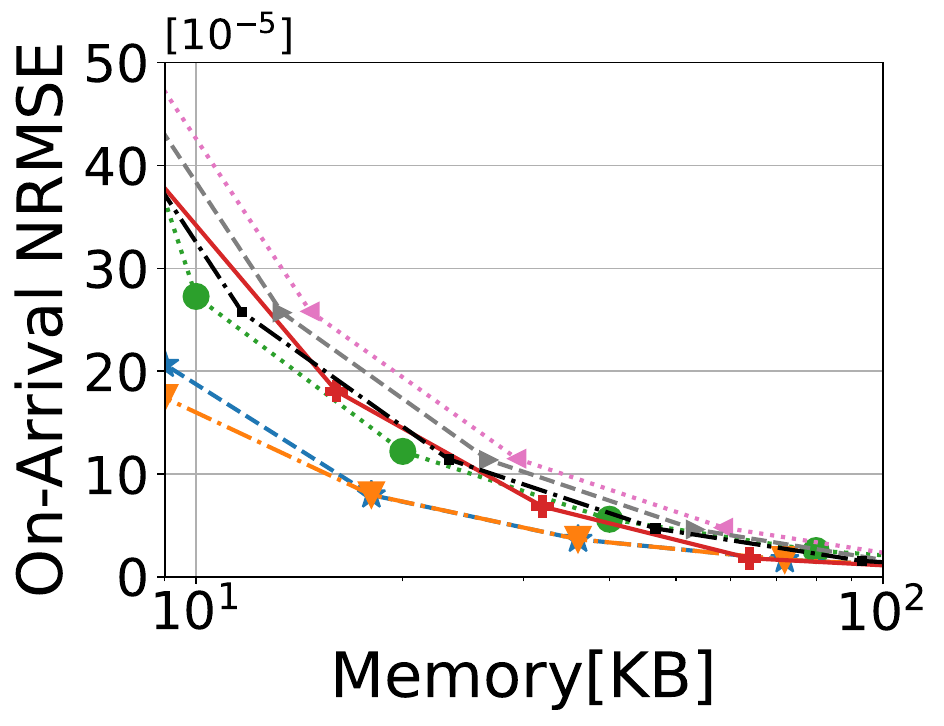}}
        \subfloat[ZIPF 1.4 (zoomed)]
        {\includegraphics[width =0.490493405\columnwidth]
        {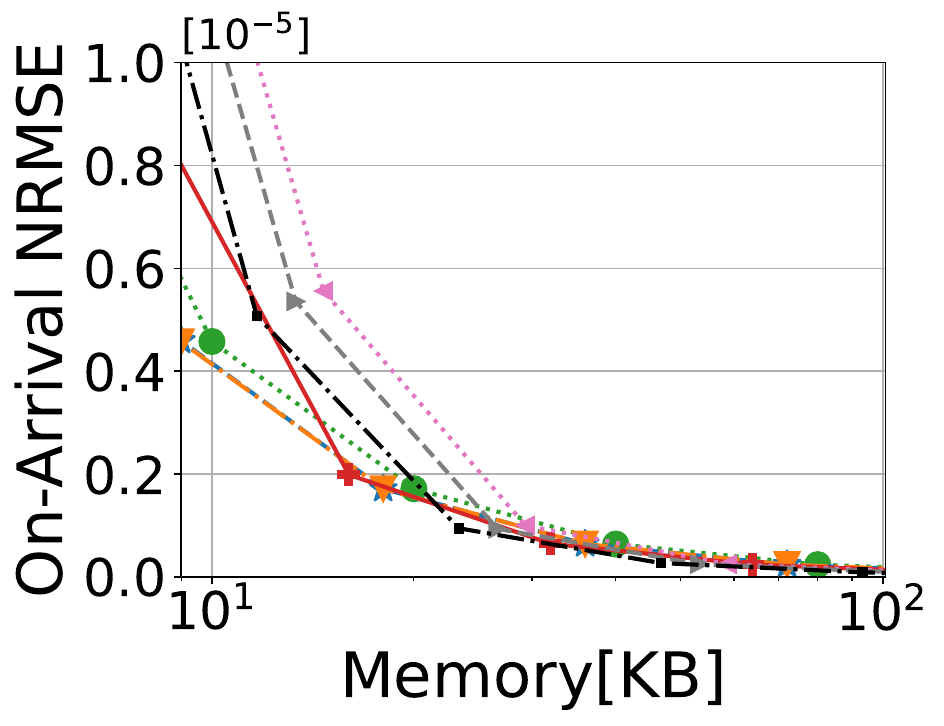}}
        
        \centering
        \vspace{1mm}
        \caption{Comparing the On Arrival Error for different Counter Pools configurations. \label{fig:conf_cmp}}
      \end{figure*}

\vspace{1mm}
\section{Applications}\label{sec:Applications}
\vspace{1mm}
While Counter Pools can be potentially applied to many data structures that employ counter arrays, here we describe two concrete applications that benefit from our design.

\vspace{1mm}
\subsection{Sketches}\label{sec:Sketches}
\vspace{1mm}
Sketches are data structures designed to measure different properties of input streams with a limited amount of space. For example, the count min sketch~\cite{CountMinSketch} allows estimating item frequencies, range queries, quantiles, and inner products. Traditionally, sketches are implemented using fixed-width (e.g., 32-bit) counters to ensure that these do not overflow when processing large data streams. However, data skew means that most counters require a lot fewer bits (see~\Cref{fig: CounterSizeCmpNYC}).
Therefore, we can use counter pools instead of individual counters \mbox{while allowing the representation of large values.}

\vspace{1mm}
\subsection{Histograms}\label{sec:Histograms}
\vspace{1mm}
Computing the histogram of an input stream is a fundamental data mining task with numerous applications. \\

The following SQL query serves as a typical example~\cite{gibbons2001distinct} where a histogram is sought:

\begin{lstlisting}[backgroundcolor = \color{lightgray!50},style=SQL, breaklines=true]
SELECT COUNT(cust_key) FROM Orders       WHERE (date >= '01-05-2025')
\end{lstlisting}

Commonly, one would compute the histogram by using a (e.g., cuckoo) hash table that maps each item into its current count; whenever an item arrives, we increment its counter if it has one and otherwise allocate it with a new counter initialized to the value of one. Similarly to the sketch use-case, common skew in data means that not all counters reach the same value and require the same number of bits. As a result, for the same memory allocation, the table will run at a higher load factor than needed, leading to slower insertion times.

We propose to compute the exact histogram of a data stream with a cuckoo hash table in which each bucket is a counter pool. Interestingly, in this application, we can completely avoid pool failures: when we need to increase the count of an item, and there are no bits available in its pool, we can migrate one of the bucket's items to its second bucket. While this means that we may require move item evictions for the same number of buckets, the space consumption of each bucket is lower due to the pool structure and we can therefore fit more buckets in the same memory space, effectively reducing the load factor and enabling tracking more items within a given space at \emph{higher} processing speeds.

   \begin{figure*}[t!]

        \subfloat[NYC 200Kb]{
        \includegraphics[width =0.24\linewidth]{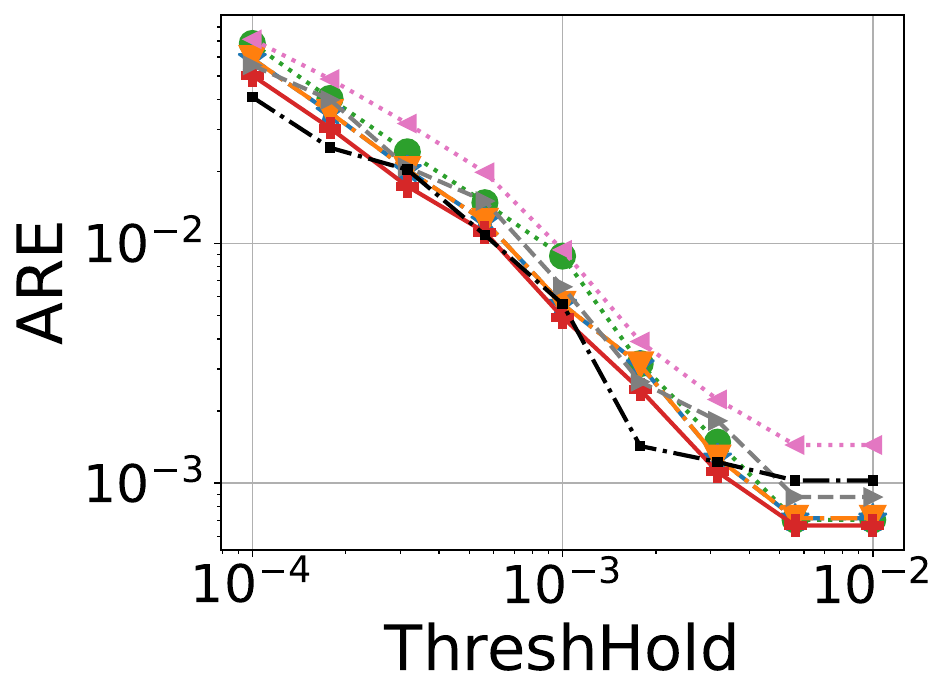}}
        \subfloat[0.6 200Kb]
        {\includegraphics[width =0.24\linewidth]
        {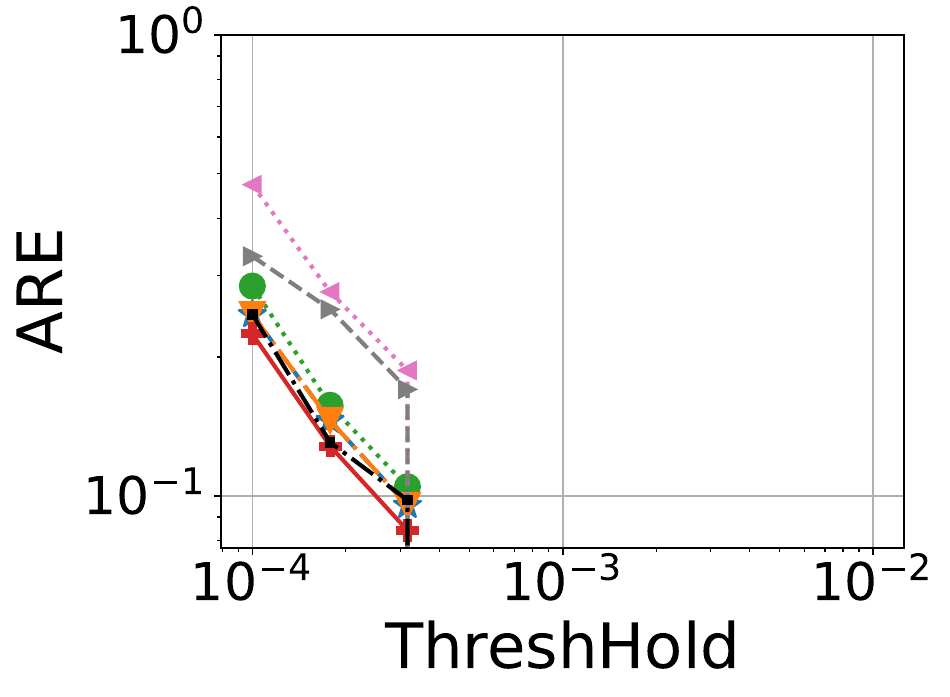}}
        \subfloat[1.0 200Kb]
        {\includegraphics[width =0.24\linewidth]{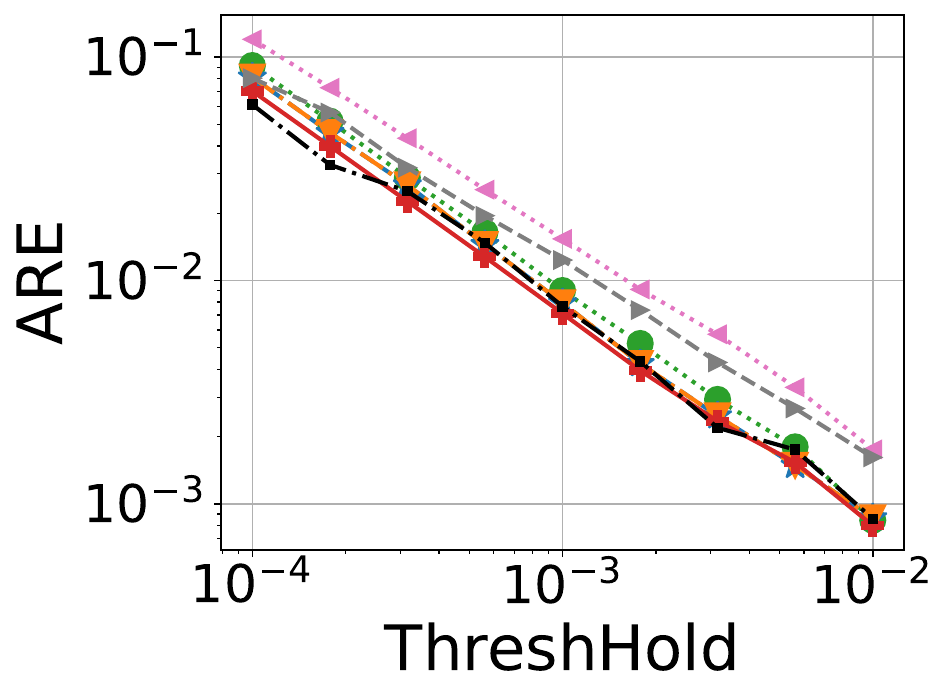}}
        \subfloat[1.4 200Kb]
        {\includegraphics[width =0.24\linewidth]
        {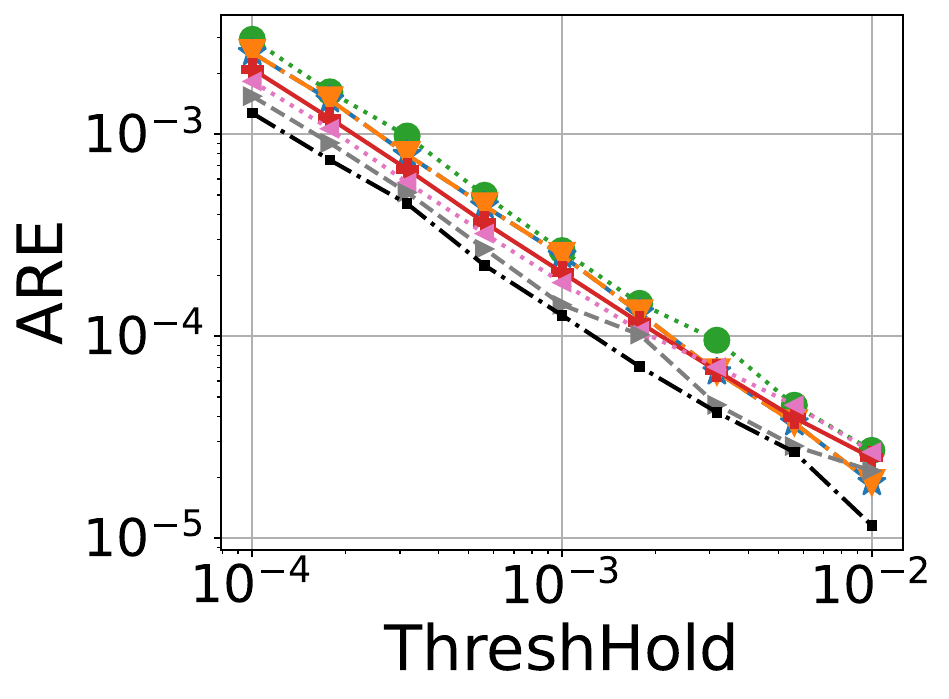}}\\
        \vspace{5mm}
                  \centering{\includegraphics[width =1\linewidth]
        {graphs/legend.pdf}}\\
        \vspace{2mm}
        \subfloat[NYC 2Mb]{
        \includegraphics[width =0.24\linewidth]{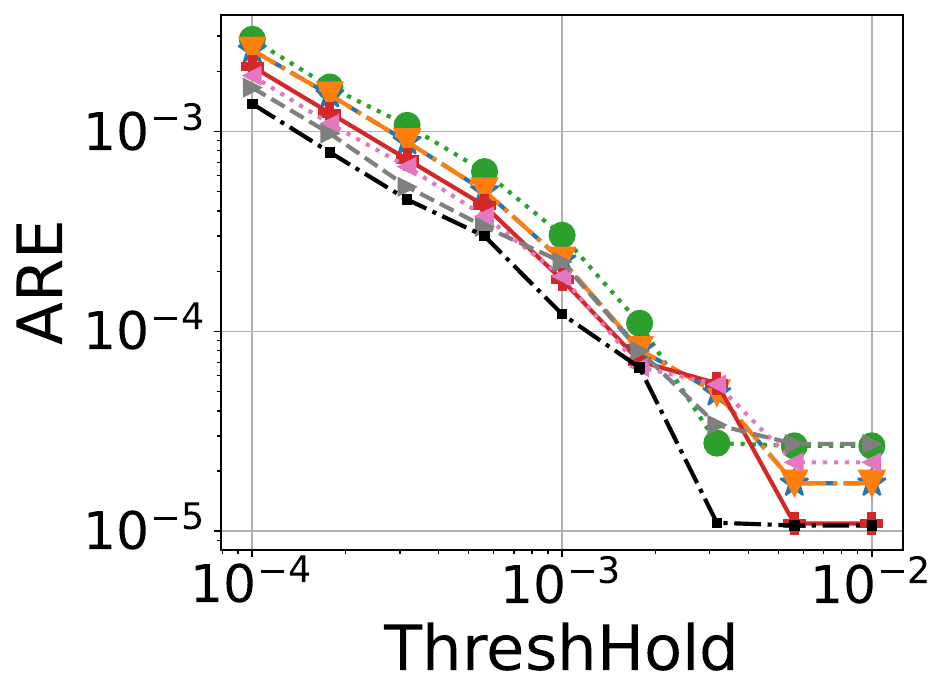}}
        \subfloat[0.6 2Mb]
        {\includegraphics[width =0.24\linewidth]
        {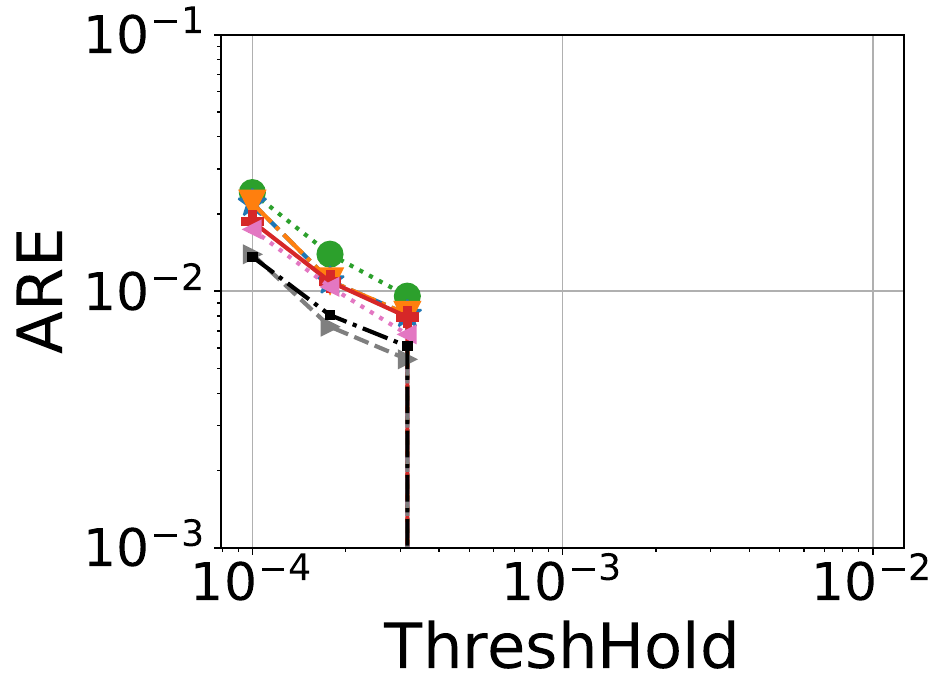}}
         \subfloat[1.0 2Mb]
        {\includegraphics[width =0.24\linewidth]{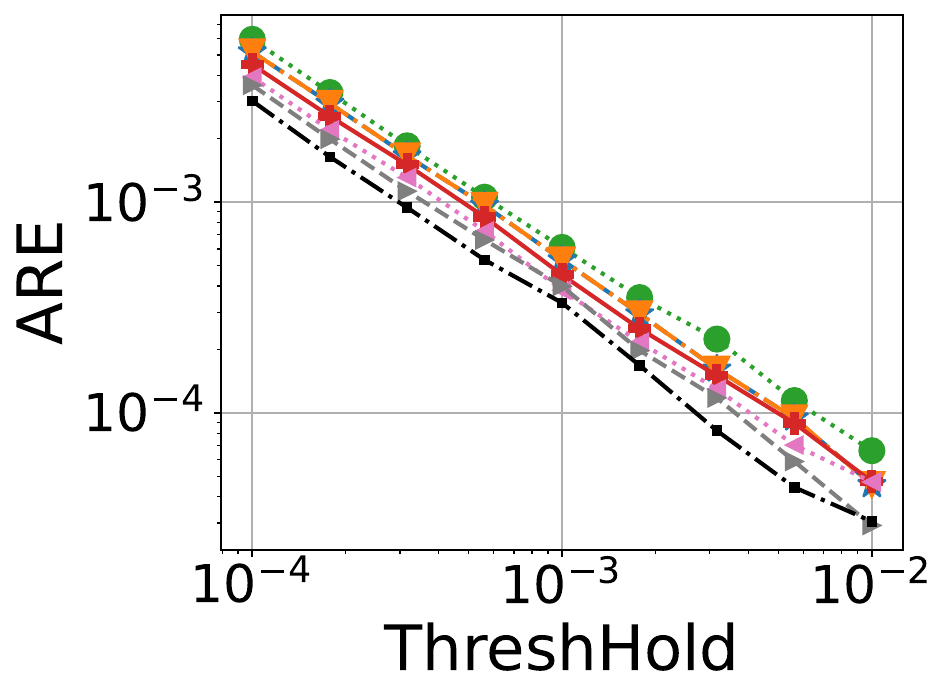}}
        \subfloat[1.4 2Mb]
        {\includegraphics[width =0.24\linewidth]{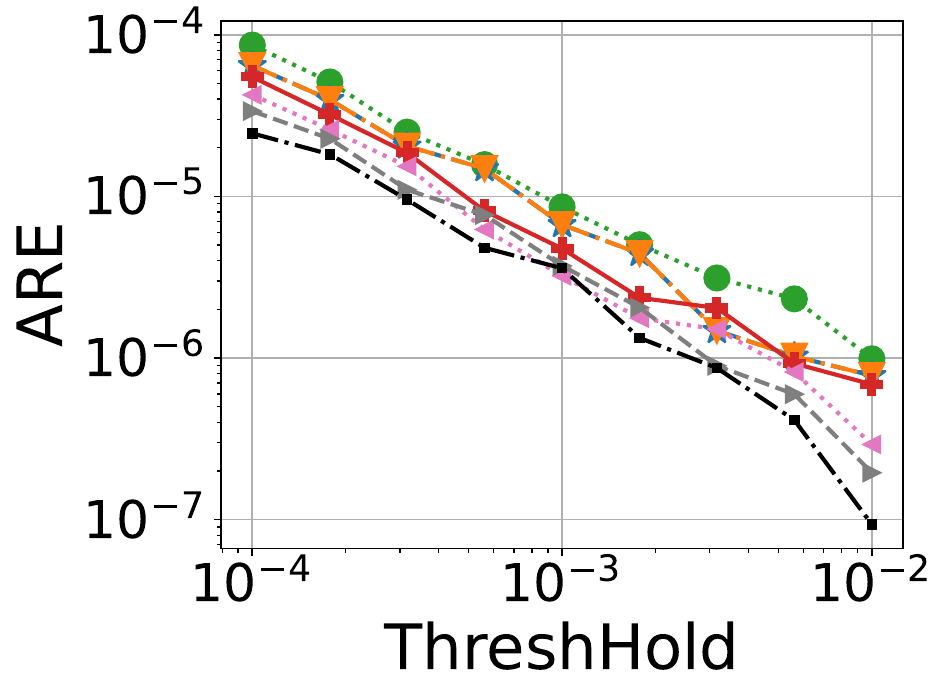}}
    \vspace{2mm}
        \caption{Comparing the ARE for heavy hitters of the different configurations with 200Kb and 2Mb memory.\label{fig:conf_cmp_HH}}
        \vspace{0mm}
      \end{figure*} 



%
%

      \begin{figure*}[t]
        \subfloat[NYC]{
        \includegraphics[width =0.25\linewidth]{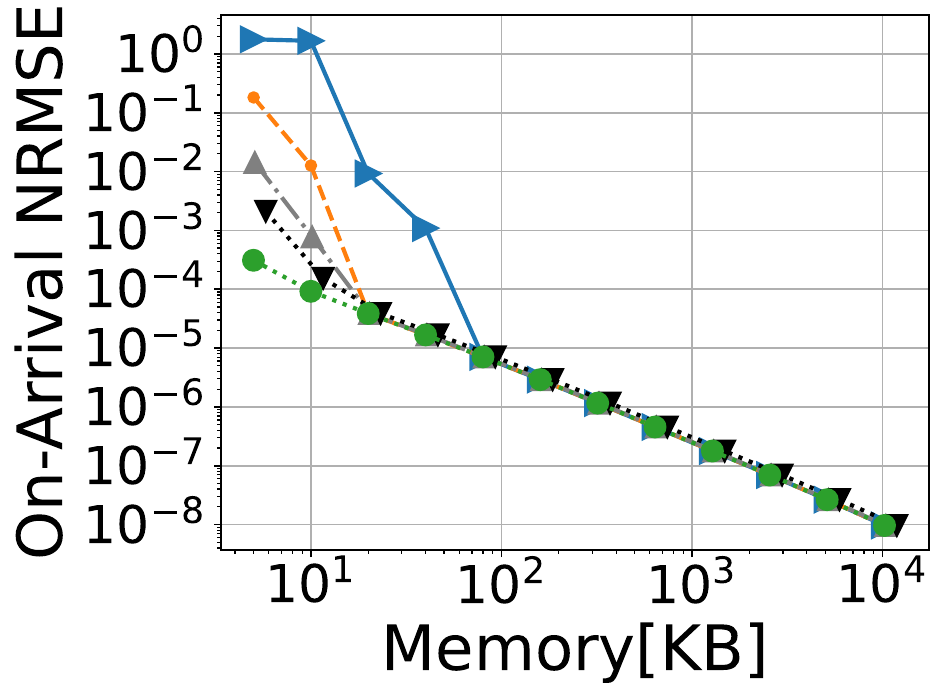}}
        \subfloat[ZIPF 0.6]{
        \includegraphics[width =0.25\linewidth]{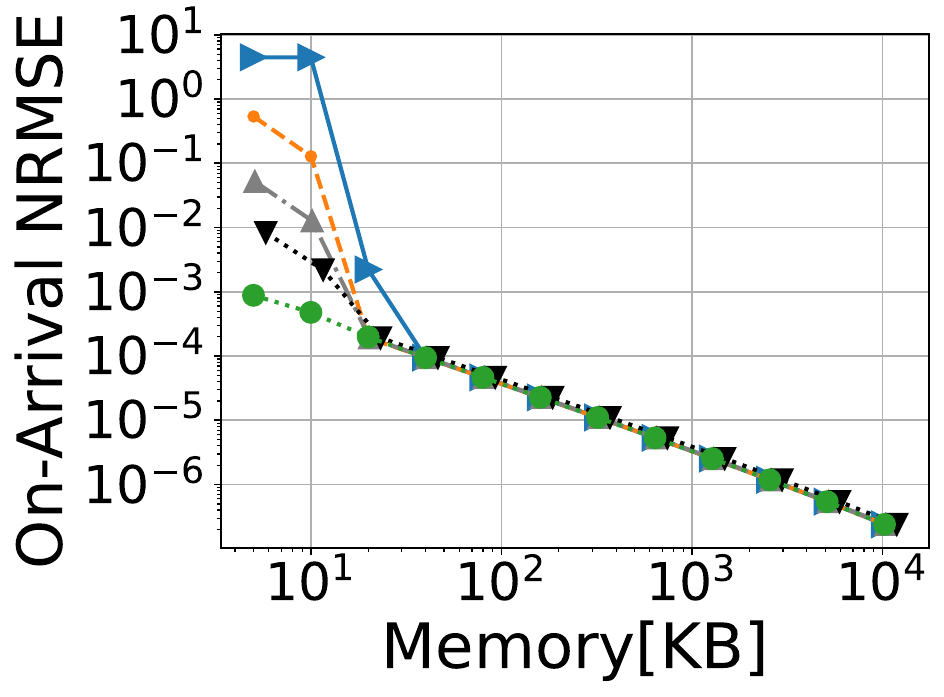}}
        \subfloat[ZIPF 1.0]{
        \includegraphics[width =0.25\linewidth]{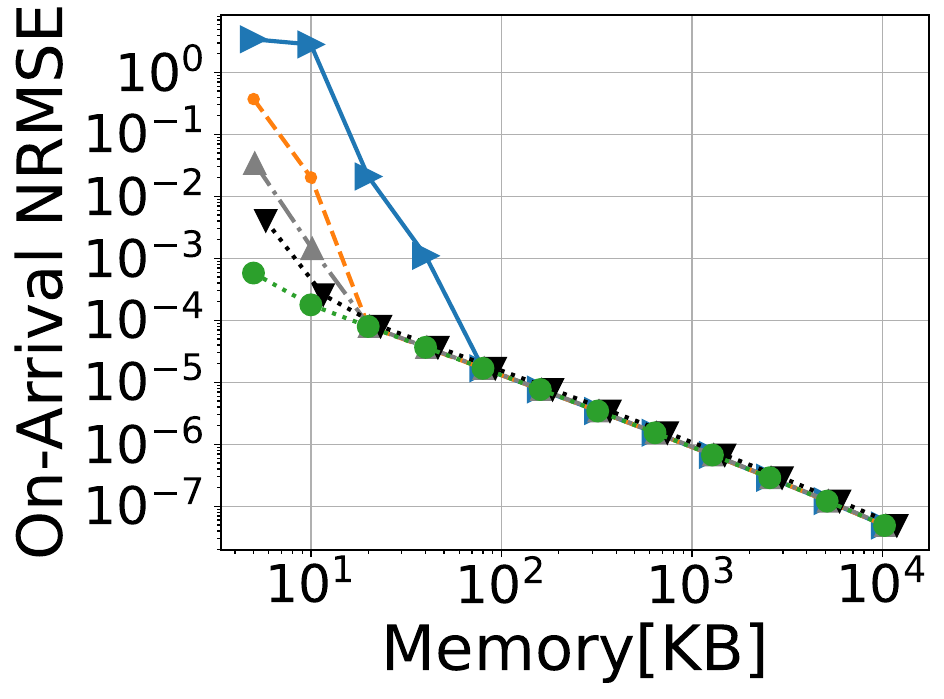}}
        \subfloat[ZIPF 1.4]{
        \includegraphics[width =0.25\linewidth]{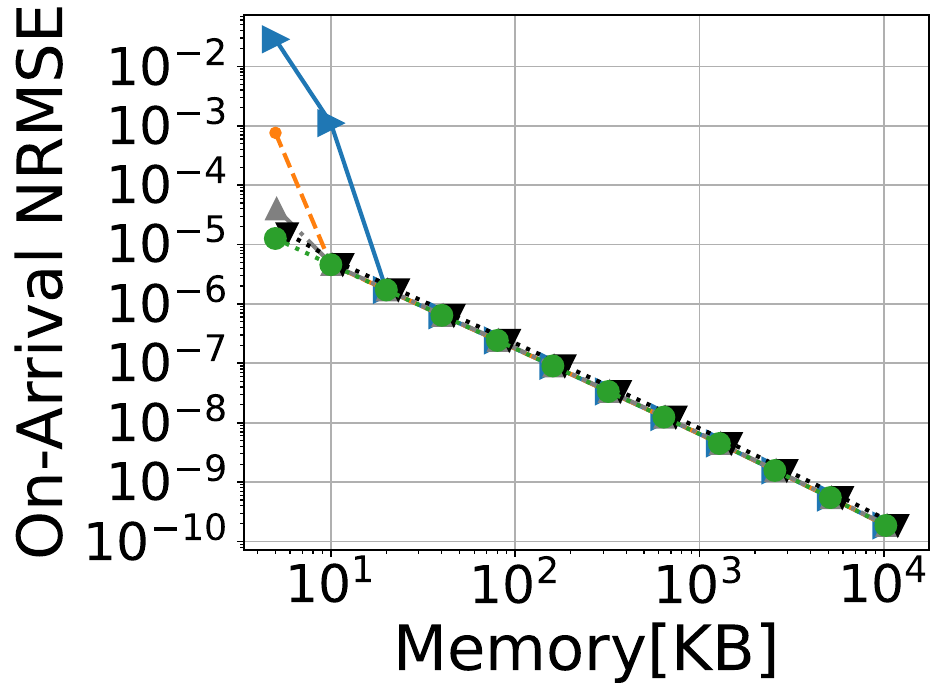}}\\
        \vspace{0mm}
         \centering{\includegraphics[width =1.0\linewidth]
        {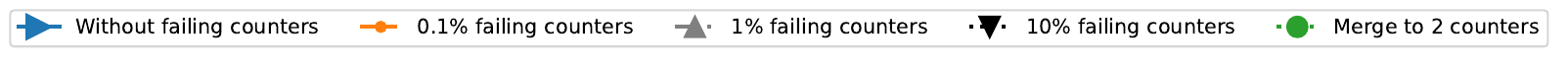}}
        \caption{Comparing different mechanisms for handling failed pools.}\label{fig: cmpFail}
\end{figure*}

        \begin{figure*}[]
        \subfloat[NYC 200Kb]{
        \includegraphics[width =0.24\linewidth]{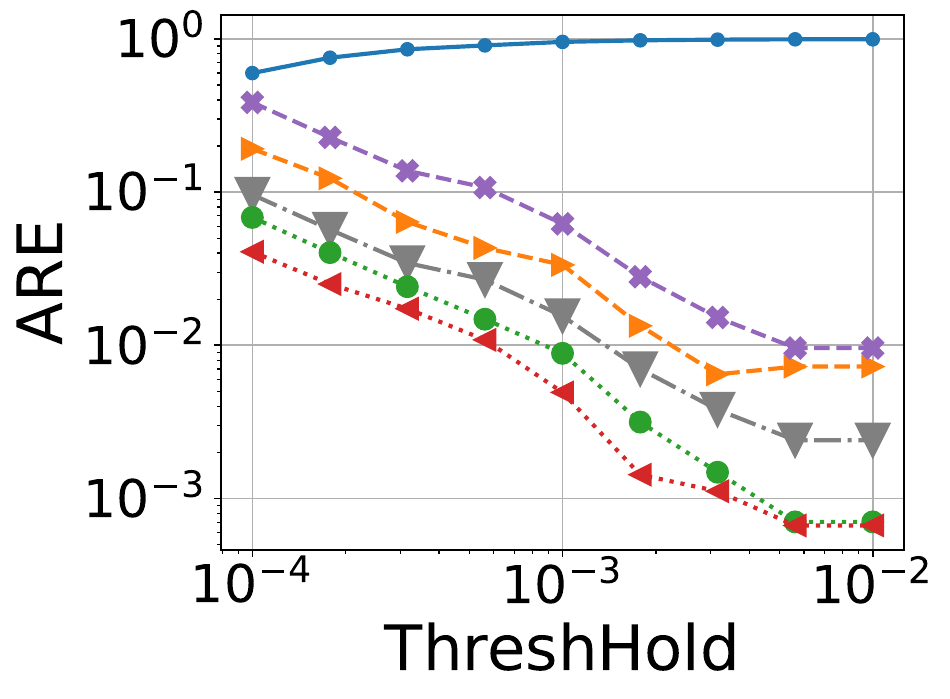}}
        \subfloat[0.6 200Kb]
        {\includegraphics[width =0.24\linewidth]
        {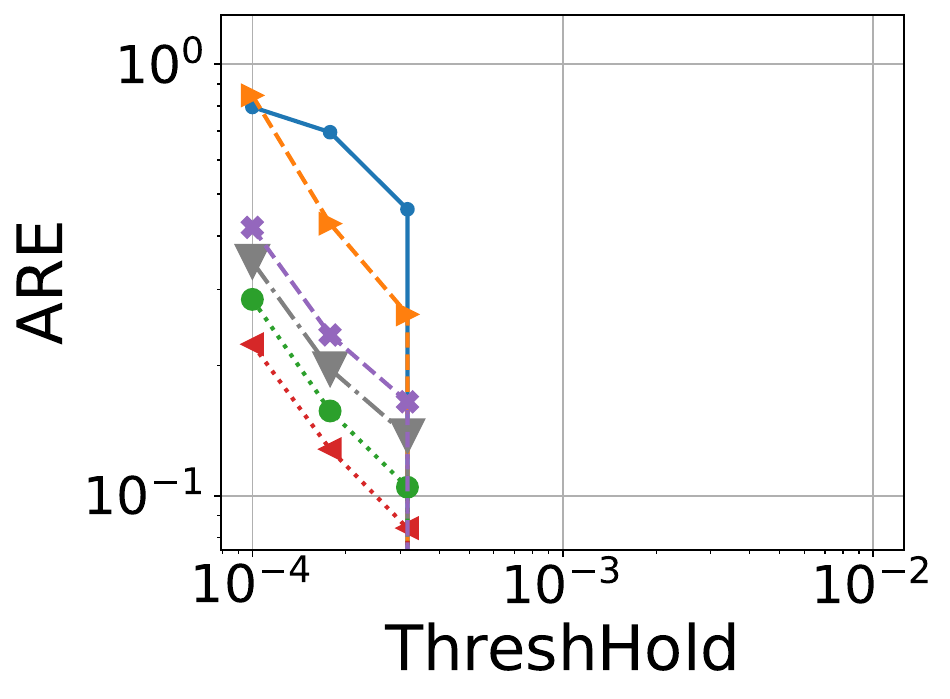}}
        \subfloat[1.0 200Kb]
        {\includegraphics[width =0.24\linewidth]{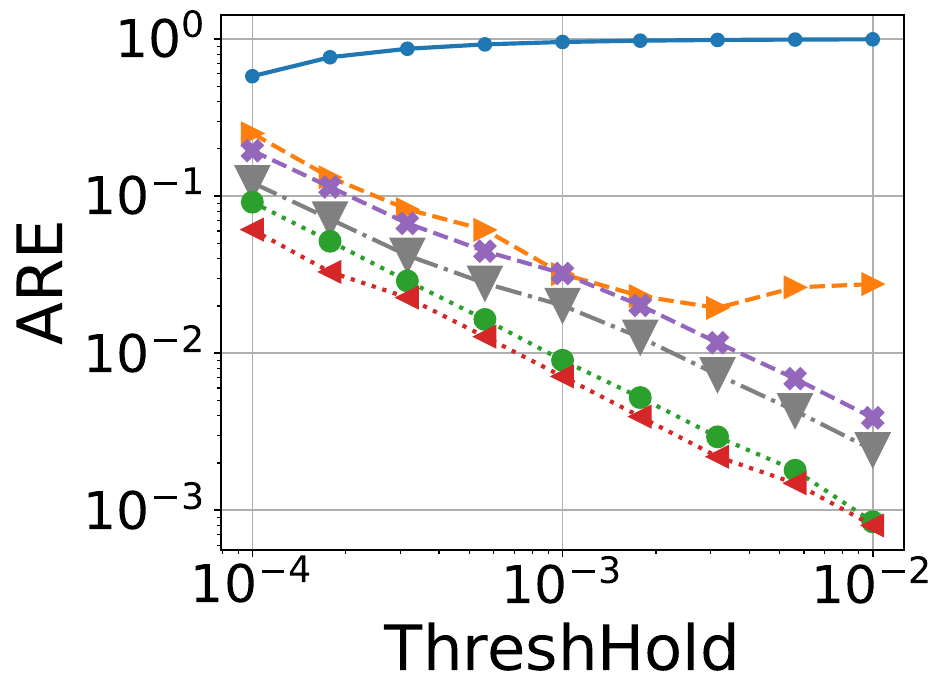}}
        \subfloat[1.4 200Kb]
        {\includegraphics[width =0.24\linewidth]
        {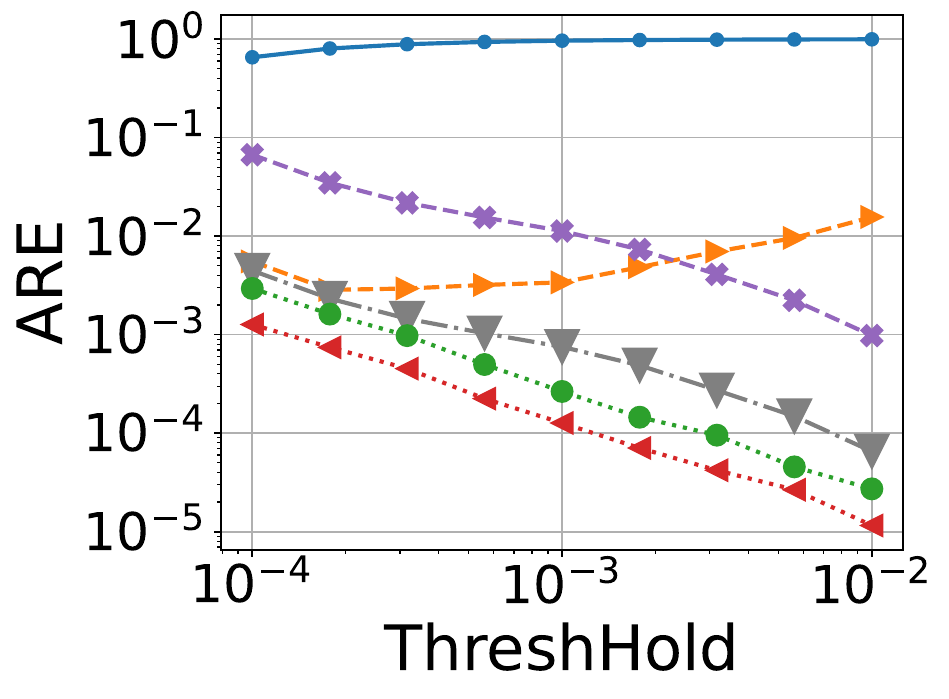}}

        {\includegraphics[width =2\columnwidth]
        {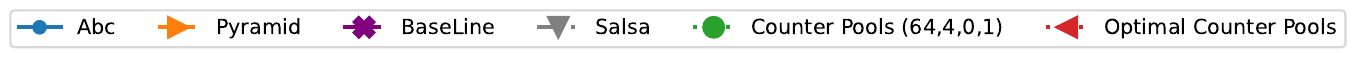}}
        \centering

        \subfloat[NYC 2Mb]{
        \includegraphics[width =0.24\linewidth]{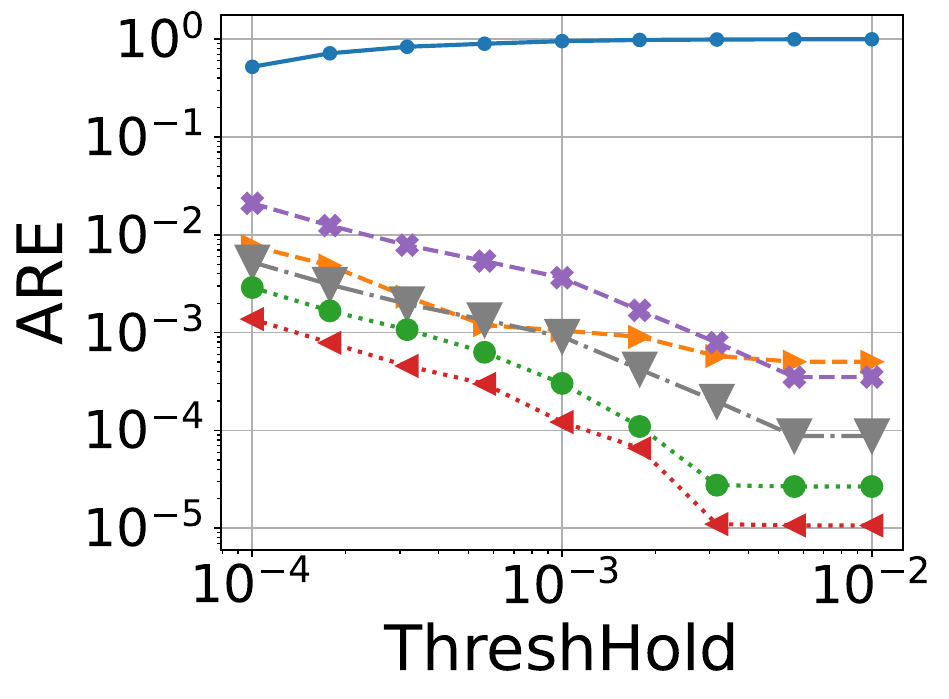}}
        \subfloat[0.6 2Mb]
        {\includegraphics[width =0.24\linewidth]
        {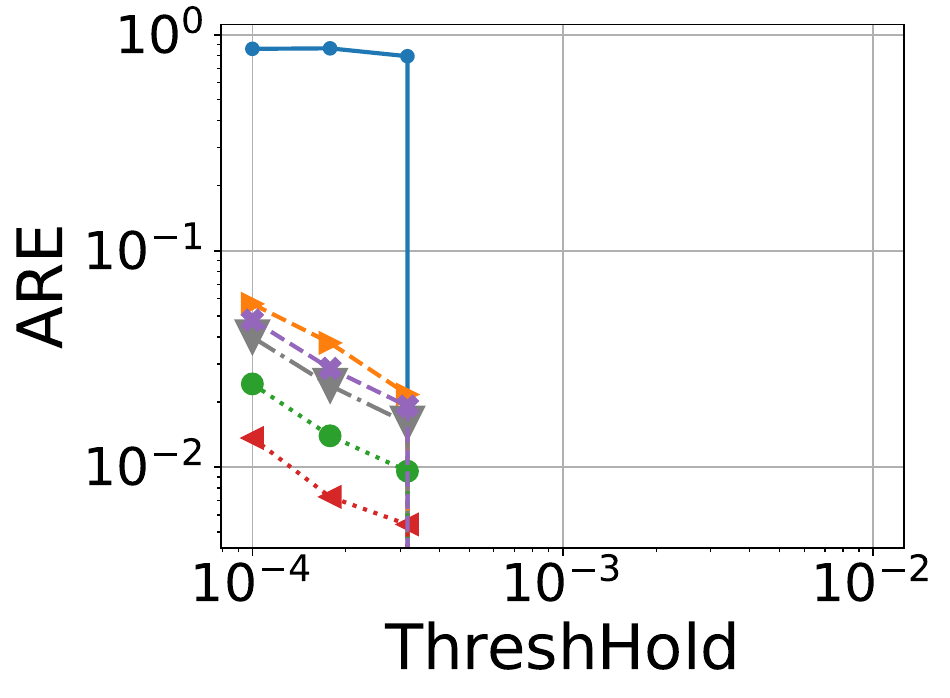}}
        \subfloat[1.0 2Mb]
        {\includegraphics[width =0.24\linewidth]{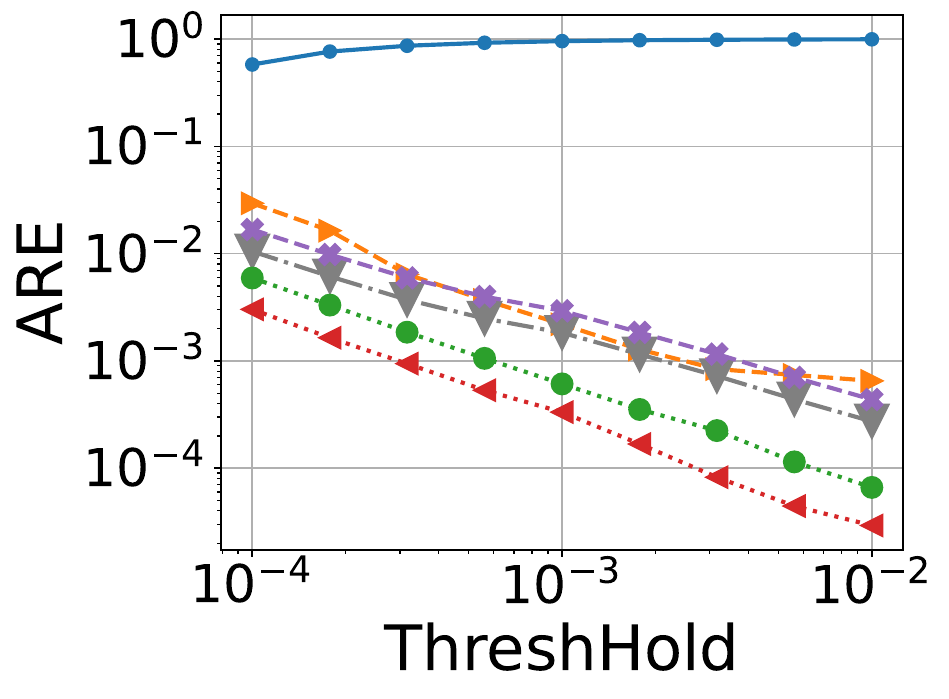}}
        \subfloat[1.4 2Mb]
        {\includegraphics[width =0.24\linewidth]
        {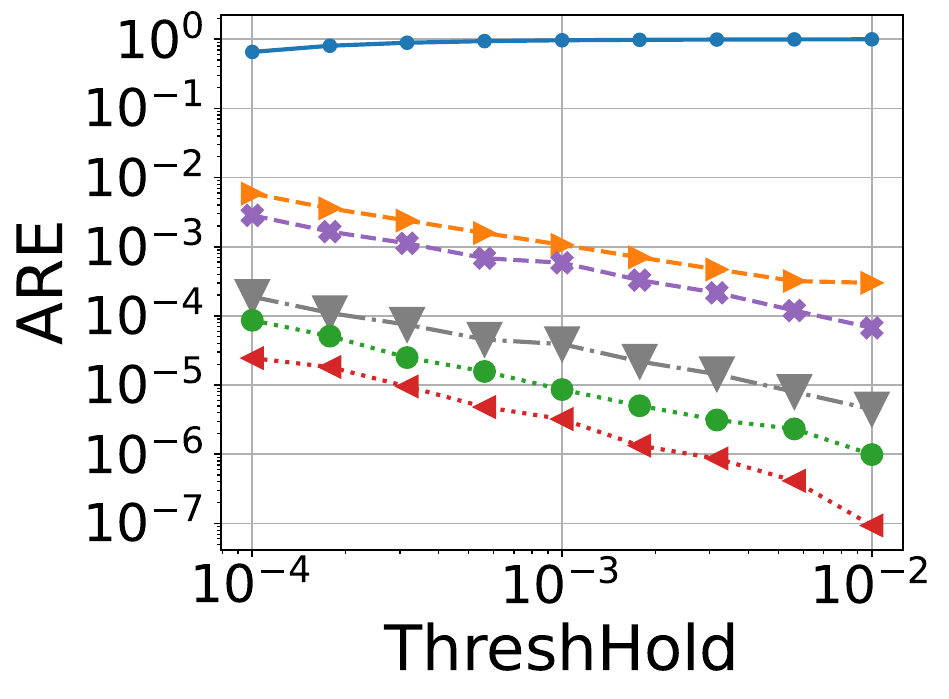}}
        
        \caption{Accuracy of heavy hitters  when varying the memory capacity.\label{fig:conf_cmp_alt_HH}}
      \end{figure*} 
       \begin{figure*}[]

        \subfloat[NYC]{
        \includegraphics[width =0.24\linewidth]{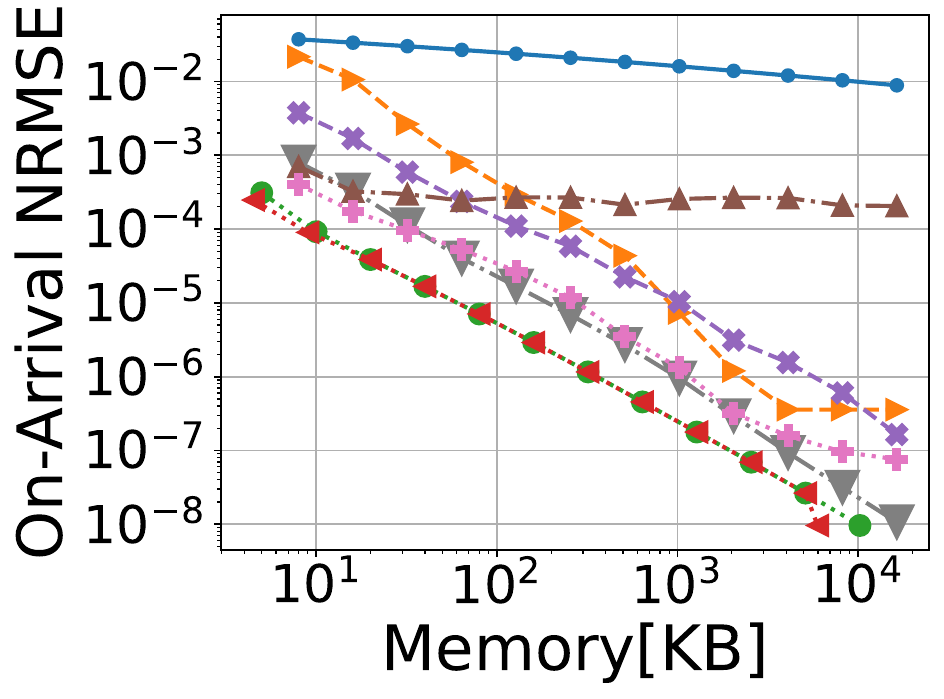}}
        \subfloat[ZIPF 0.6]
        {\includegraphics[width =0.24\linewidth]
        {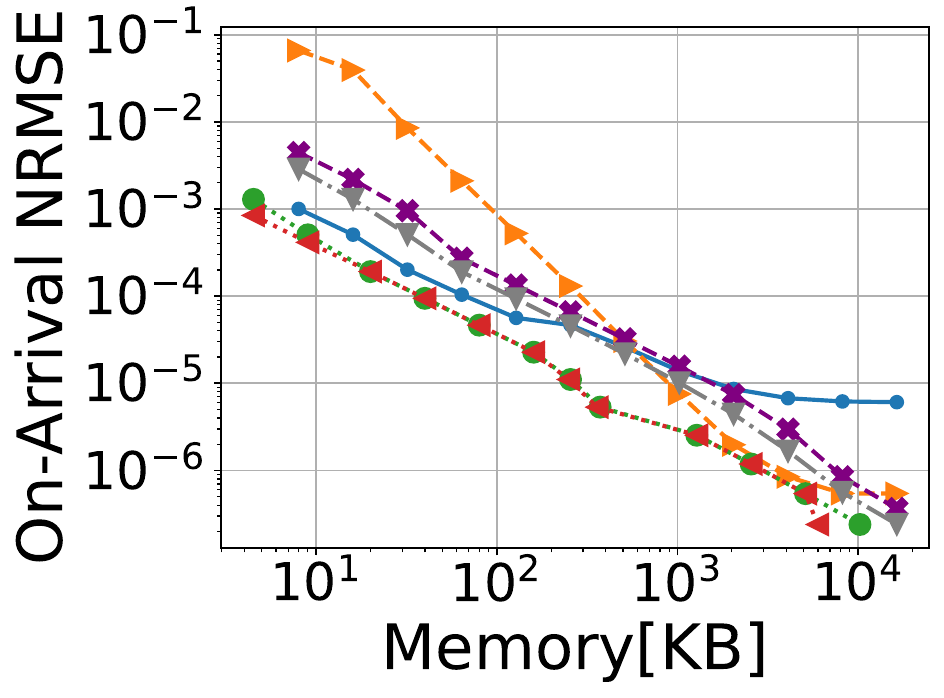}}
        \subfloat[ZIPF 1.0]
        {\includegraphics[width =0.24\linewidth]{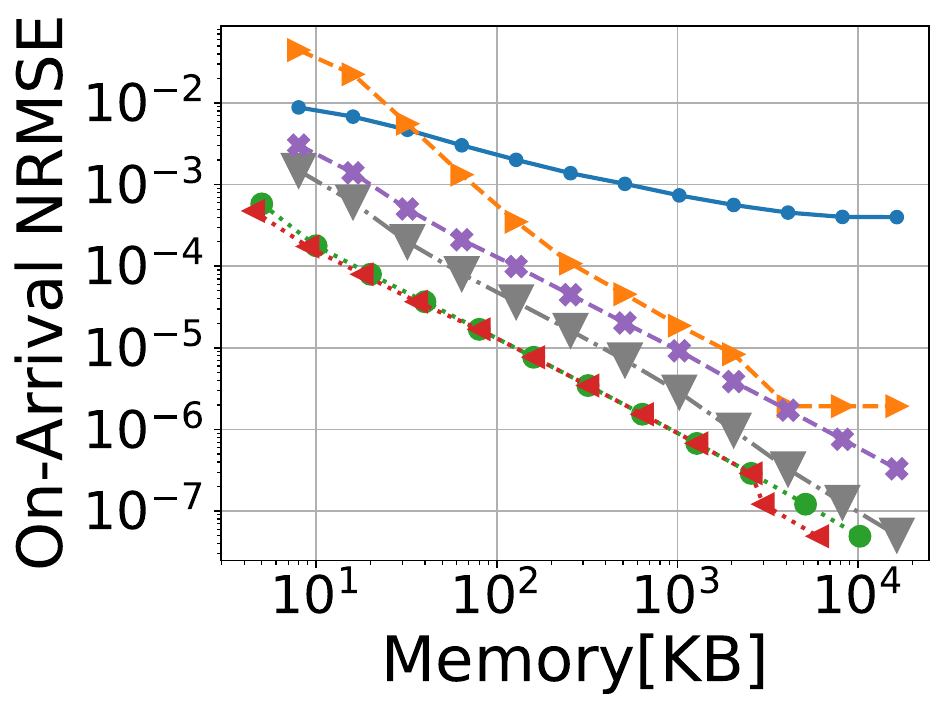}}
        \subfloat[ZIPF 1.4]
        {\includegraphics[width =0.24\linewidth]    
        {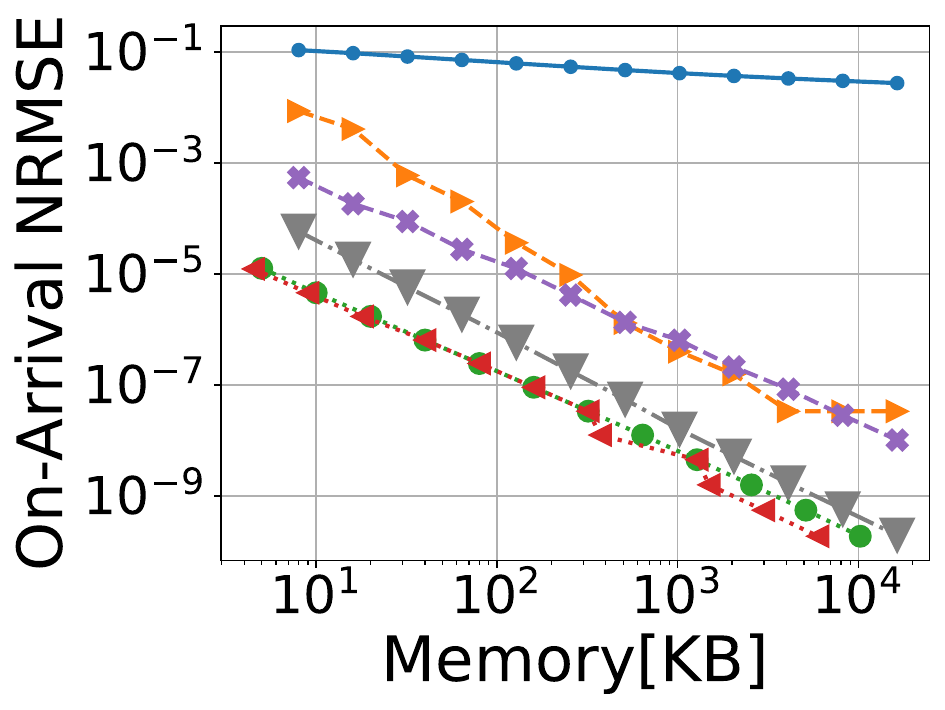}}
        
 \centering{\includegraphics[width =2\columnwidth]
        {graphs/legend2.pdf}}

        \subfloat[NYC]{
        \includegraphics[width =0.24\linewidth]{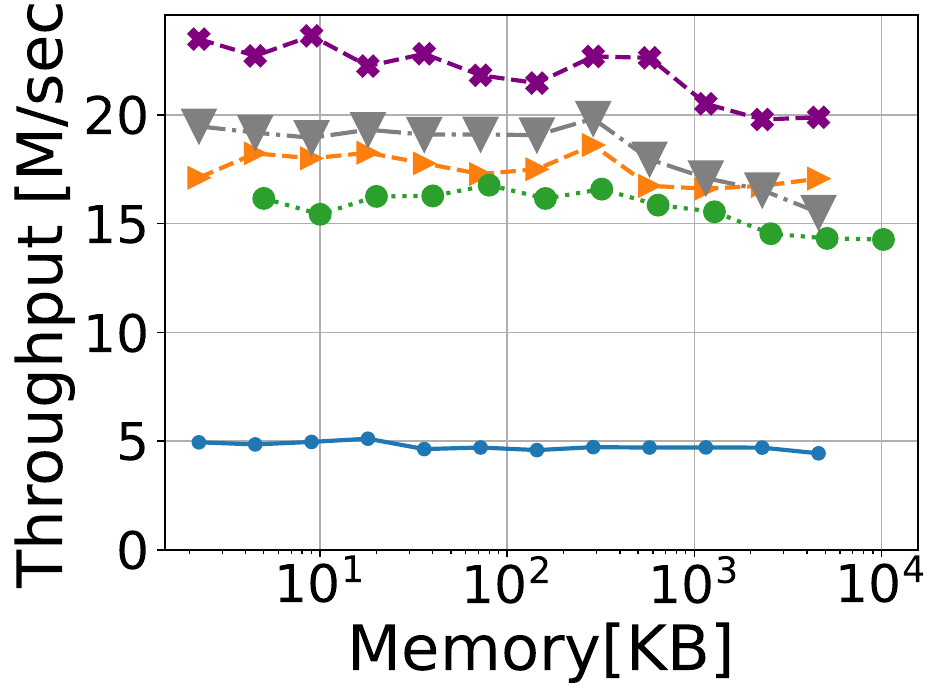}}
        \subfloat[ZIPF 0.6]
        {\includegraphics[width =0.24\linewidth]
        {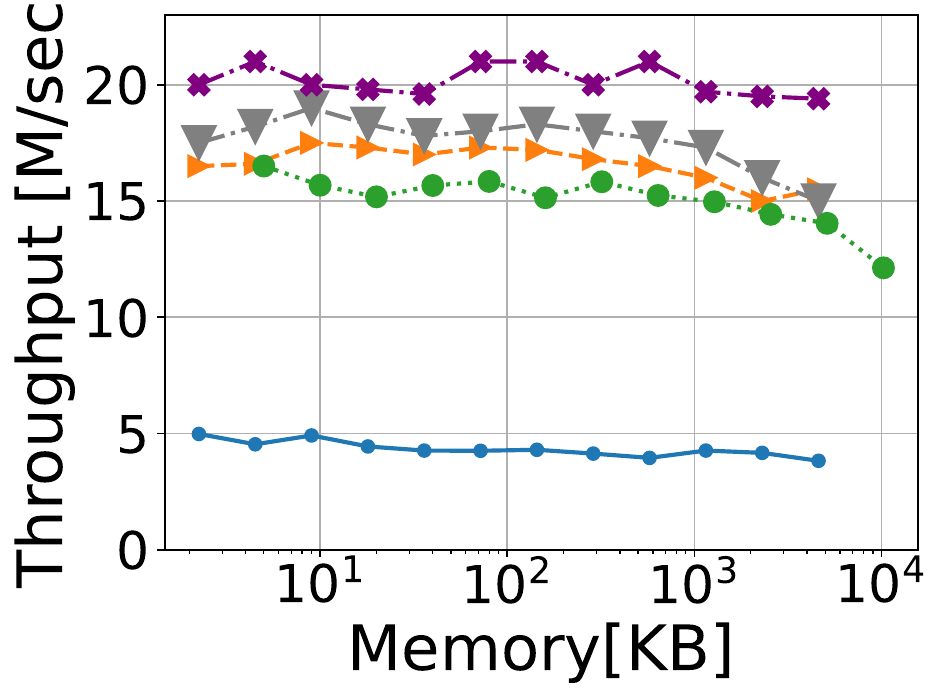}} 
        \subfloat[ZIPF 1.0]
        {\includegraphics[width =0.24\linewidth]{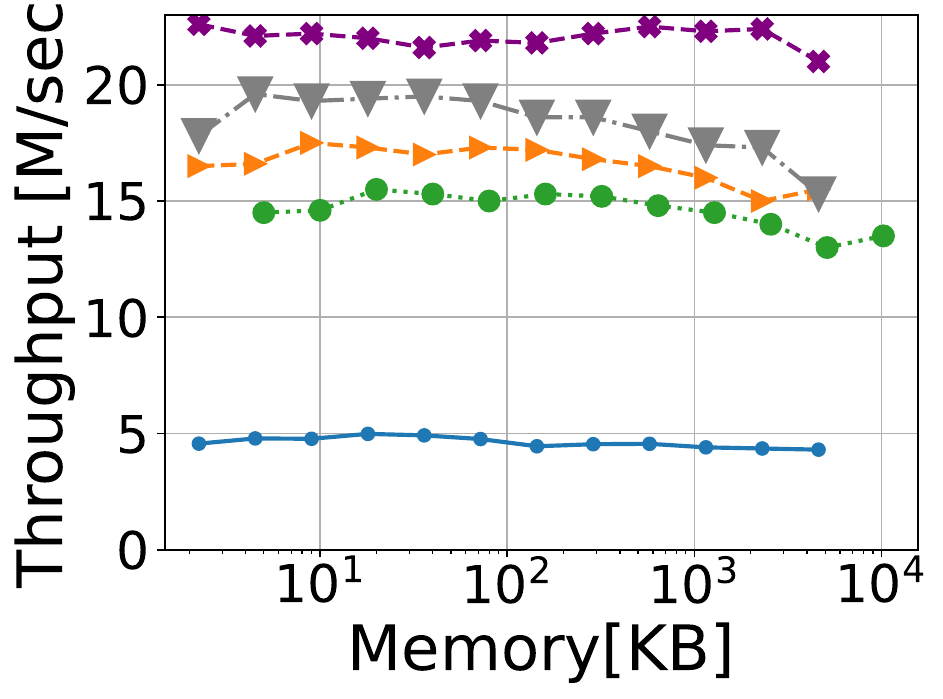}}
        \subfloat[ZIPF 1.4]
        {\includegraphics[width =0.24\linewidth]    
        {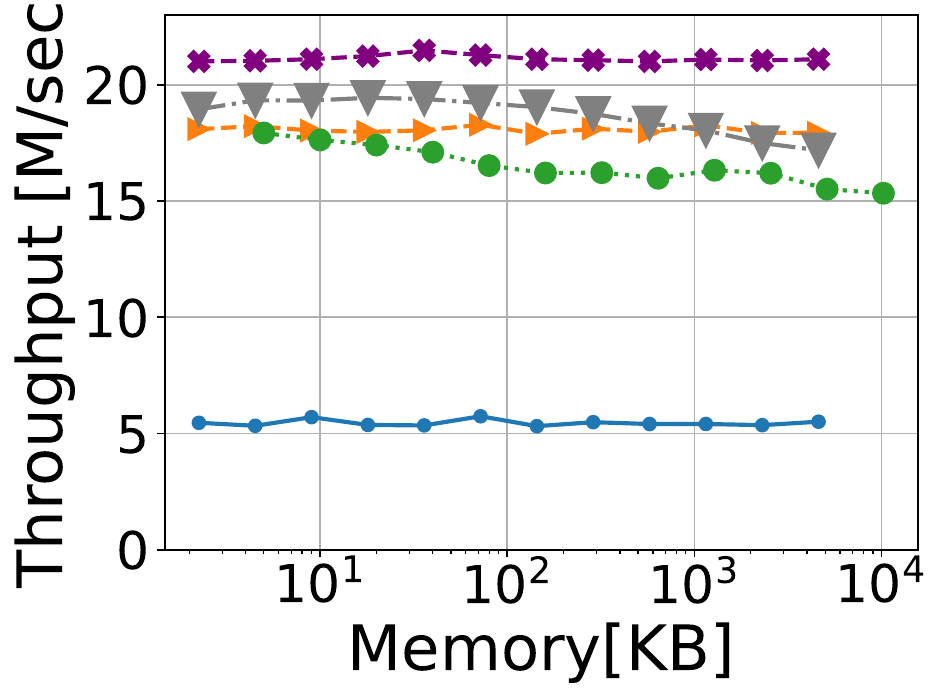}}
        
         \centering
            \caption{Comparing throughput and On-Arrival NRMSE of Counter Pools with baselines, ABC, \mbox{Pyramid Sketch and SALSA.\label{fig:cmp_speed_OnArrival}}}
      \end{figure*}

      \begin{figure*}[]

        \subfloat[NYC]{
        \includegraphics[width =0.24\linewidth]{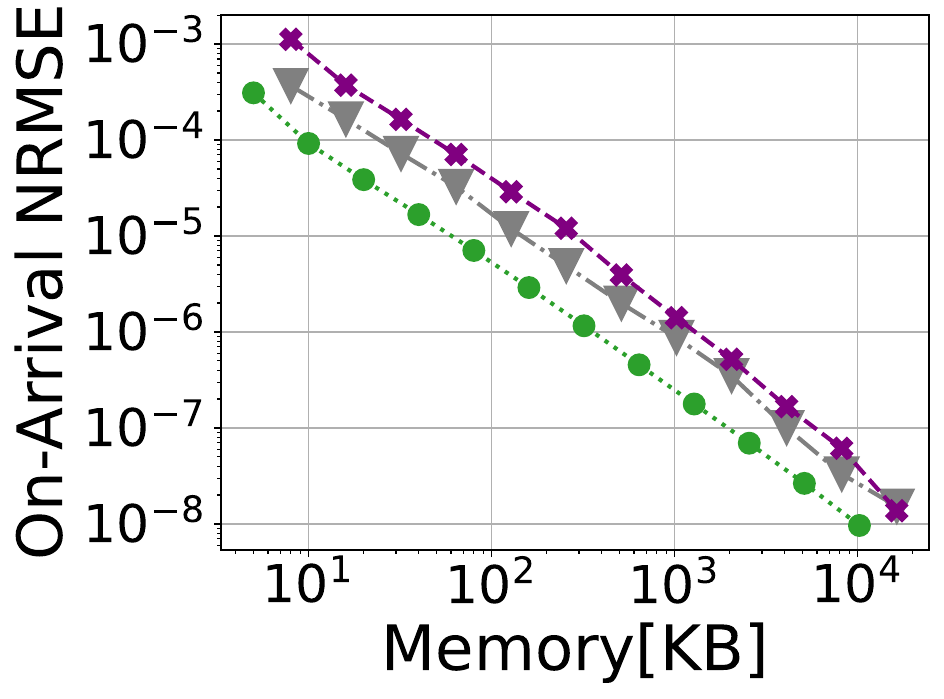}}
        \subfloat[ZIPF 0.6]
        {\includegraphics[width =0.24\linewidth]
        {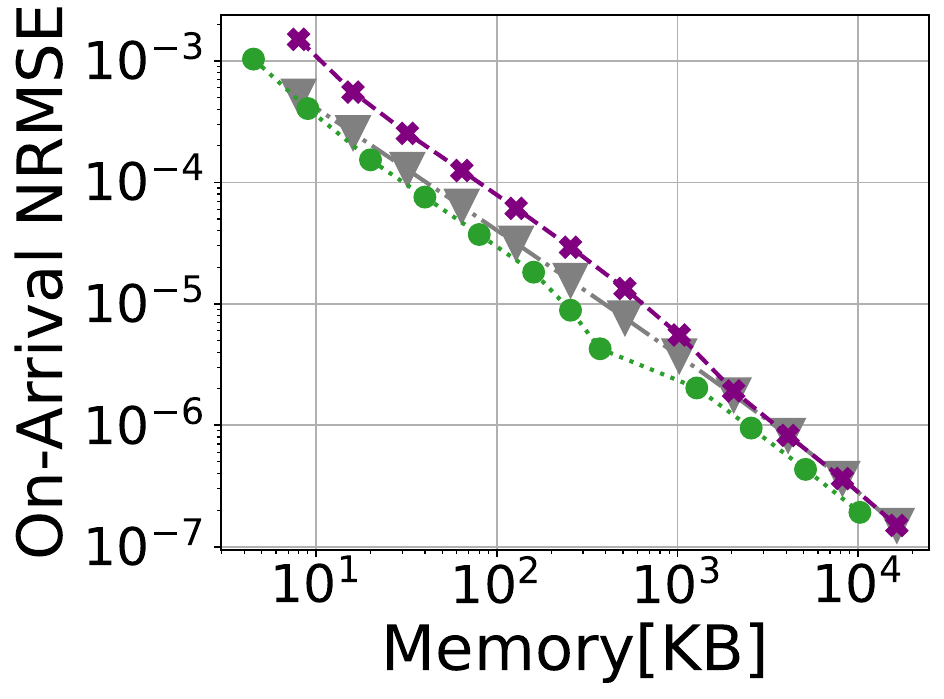}}
        \subfloat[ZIPF 1.0]
        {\includegraphics[width =0.24\linewidth]{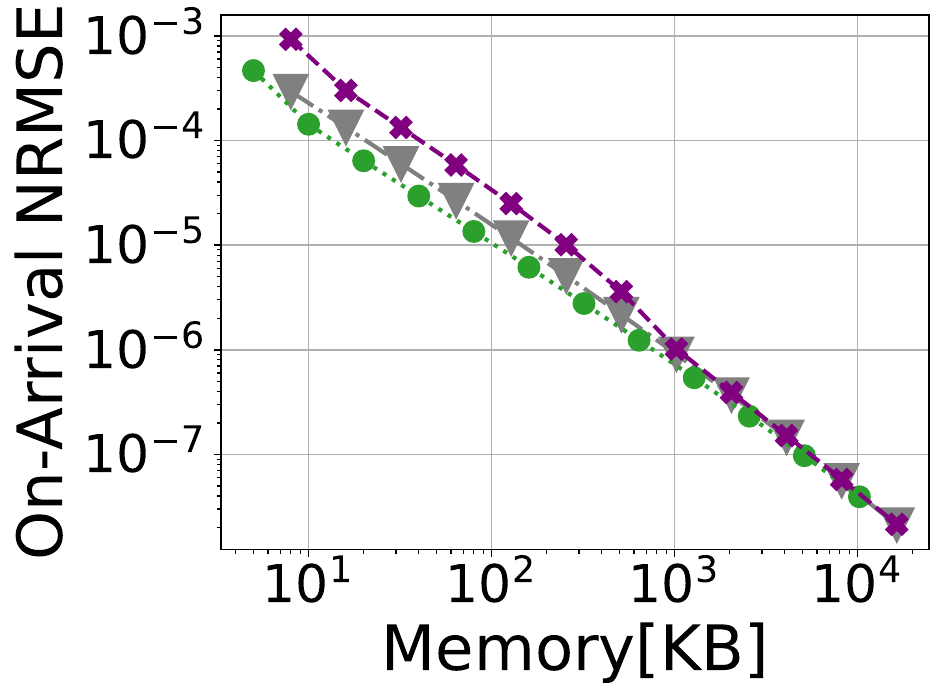}}
        \subfloat[ZIPF 1.4]
        {\includegraphics[width =0.24\linewidth]    
        {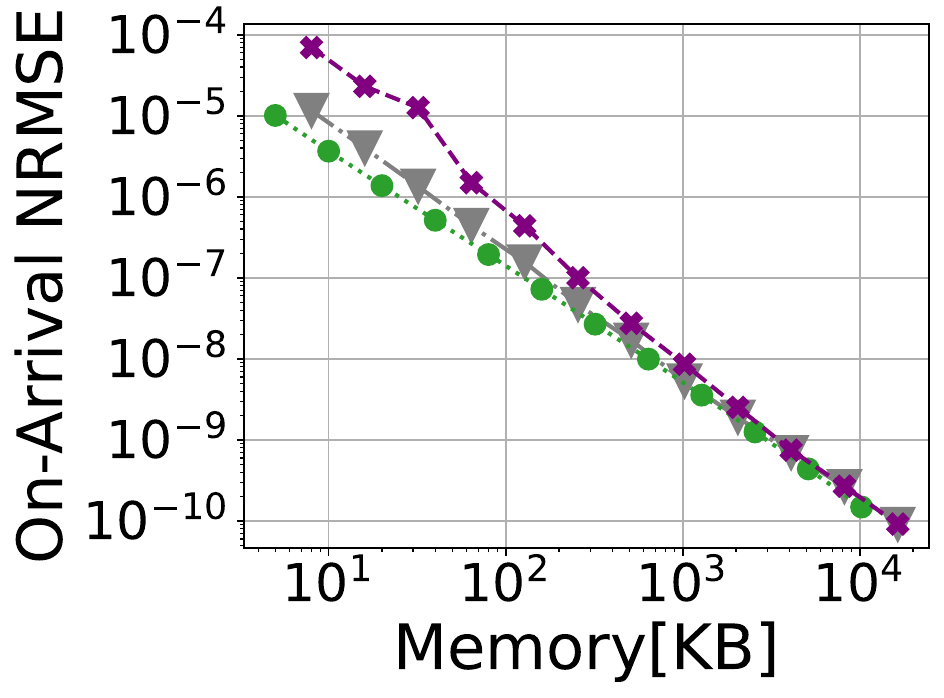}}
        \vspace{3mm}
 \centering{\includegraphics[width =1.1\columnwidth]
        {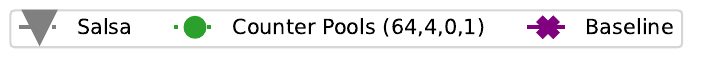}}
               \vspace{-0.2cm}

        \subfloat[NYC]{
        \includegraphics[width =0.24\linewidth]{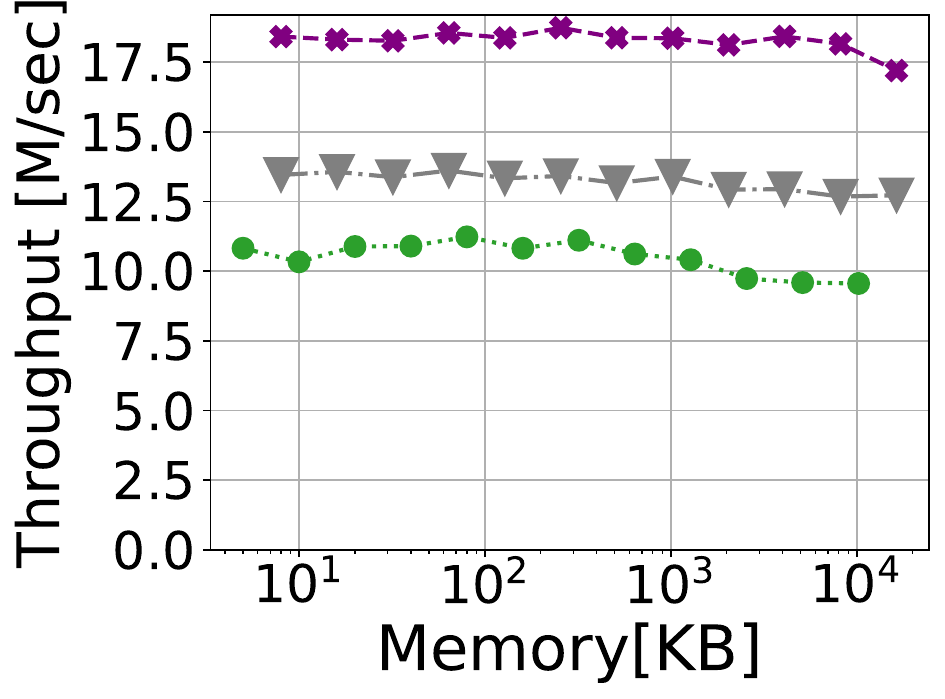}}
        \subfloat[ZIPF 0.6]
        {\includegraphics[width =0.24\linewidth]
        {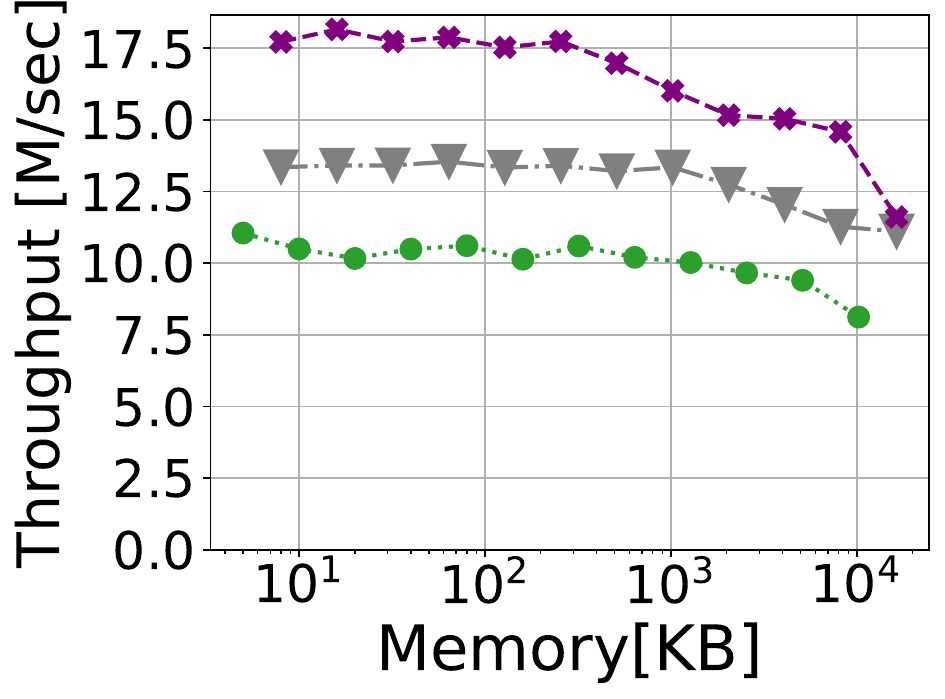}} 
        \subfloat[ZIPF 1.0]
        {\includegraphics[width =0.24\linewidth]{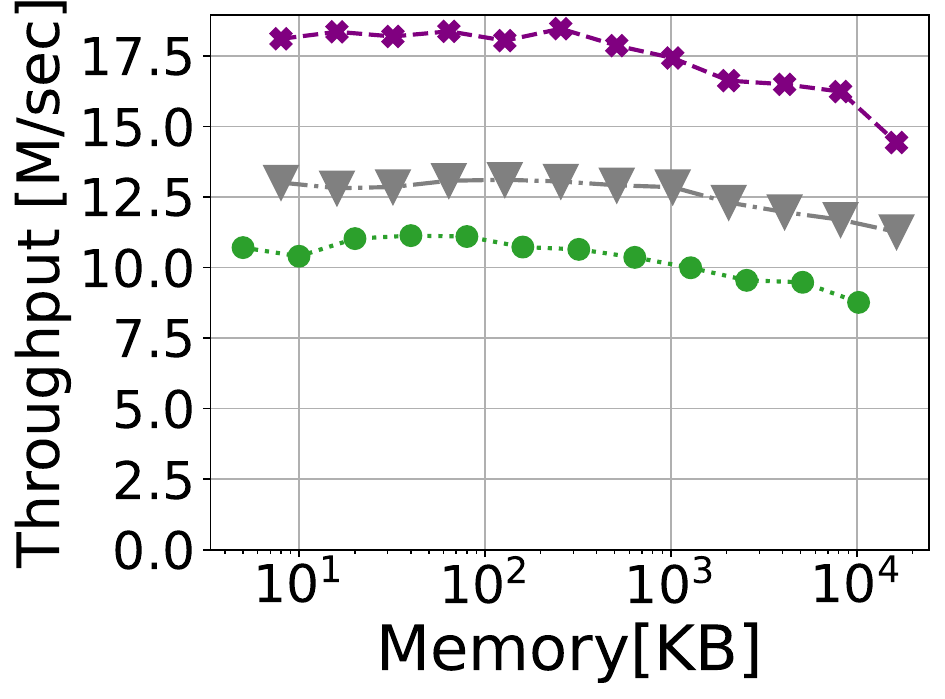}}
        \subfloat[ZIPF 1.4]
        {\includegraphics[width =0.24\linewidth]    
        {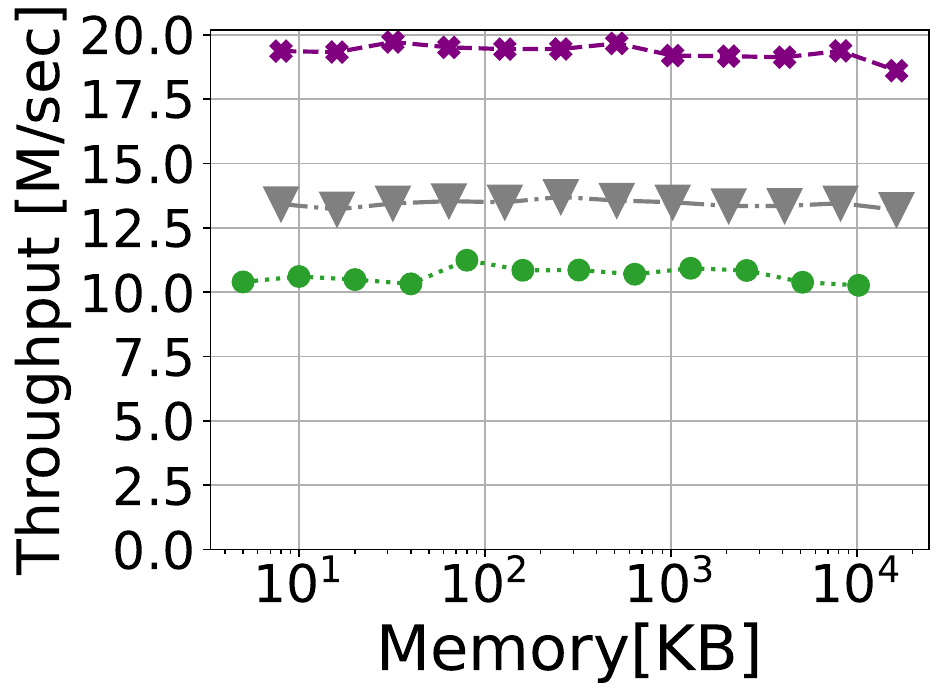}}
        
         \centering
            \caption{Comparing throughput and On-Arrival NRMSE of the Conservative Update variants of our Counter Pools, the baseline and SALSA.\label{fig:cus_speed_OnArrival}}
      \end{figure*}

           \begin{figure*}[t]
        \subfloat[NYC]{
        \includegraphics[width =0.66\columnwidth]{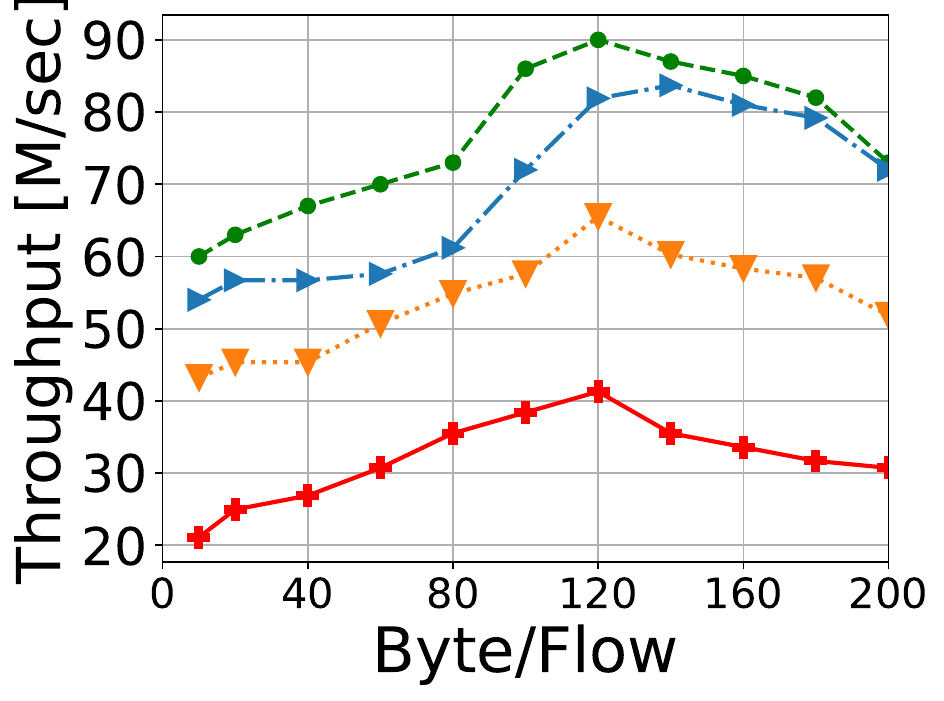}}
        \subfloat[ZIPF 0.6]
        {\includegraphics[width =0.66\columnwidth]
        {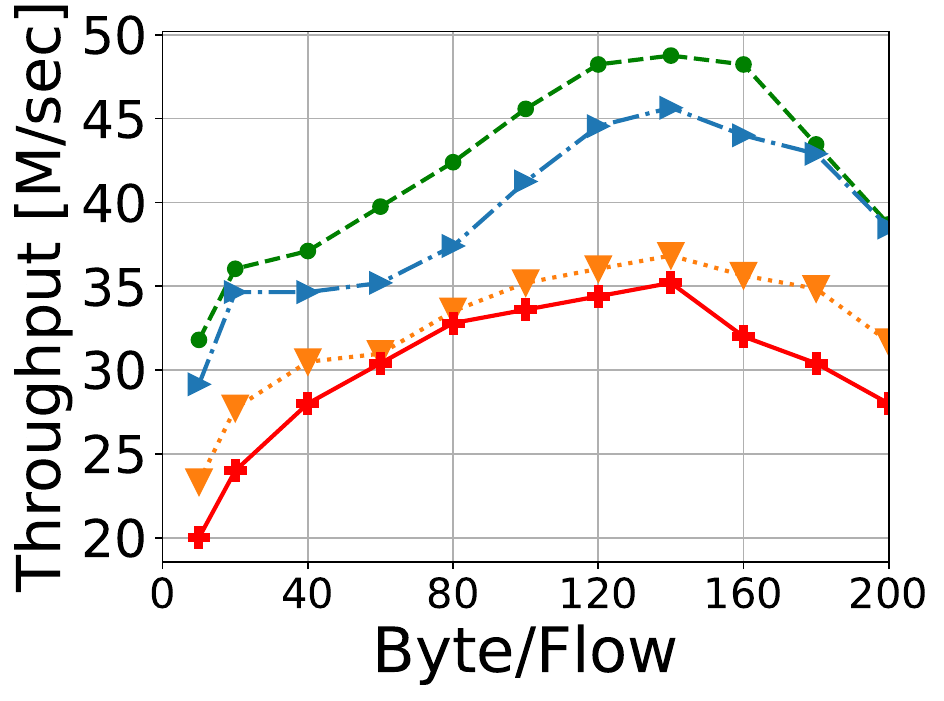}}
        \subfloat[ZIPF 0.8]
        {\includegraphics[width =0.66\columnwidth]{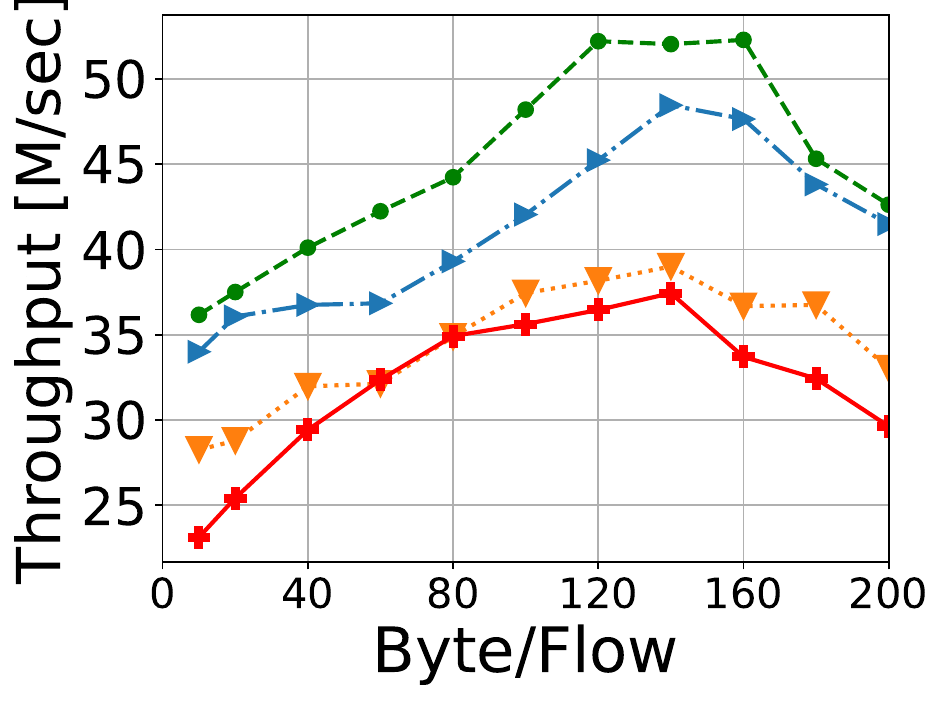}}
        \\
        \vspace{3mm}
        {\includegraphics[width =1.6\columnwidth]
        {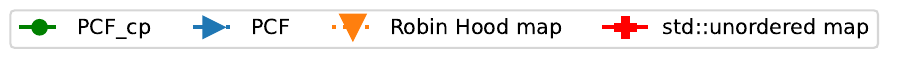}}
        \centering
        \subfloat[ZIPF 1.0]
        {\includegraphics[width =0.66\columnwidth]{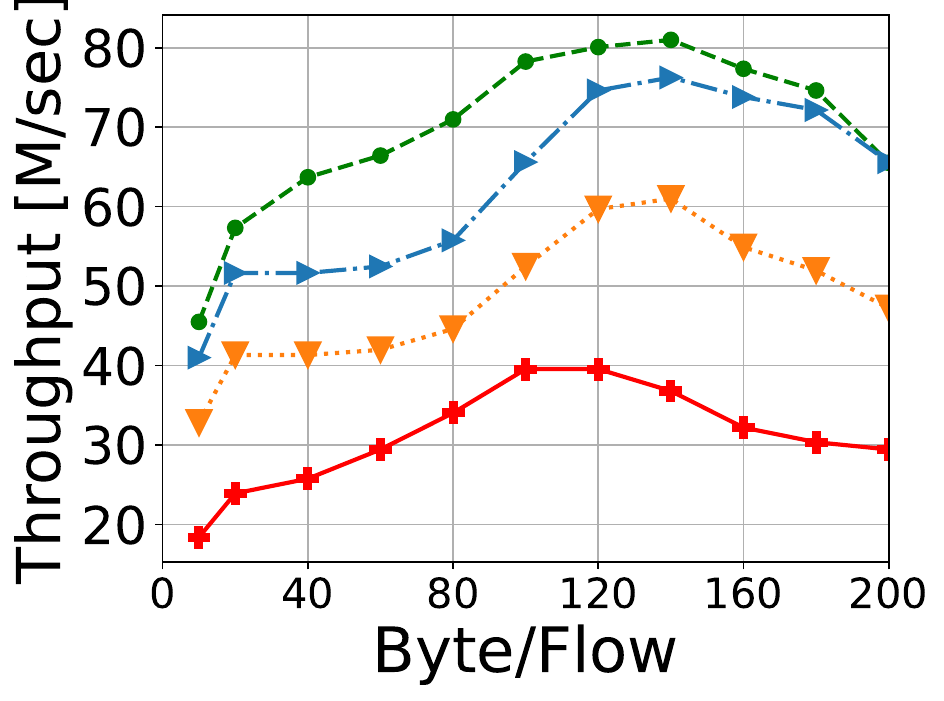}}
        \subfloat[ZIPF 1.2]
        {\includegraphics[width =0.66\columnwidth]{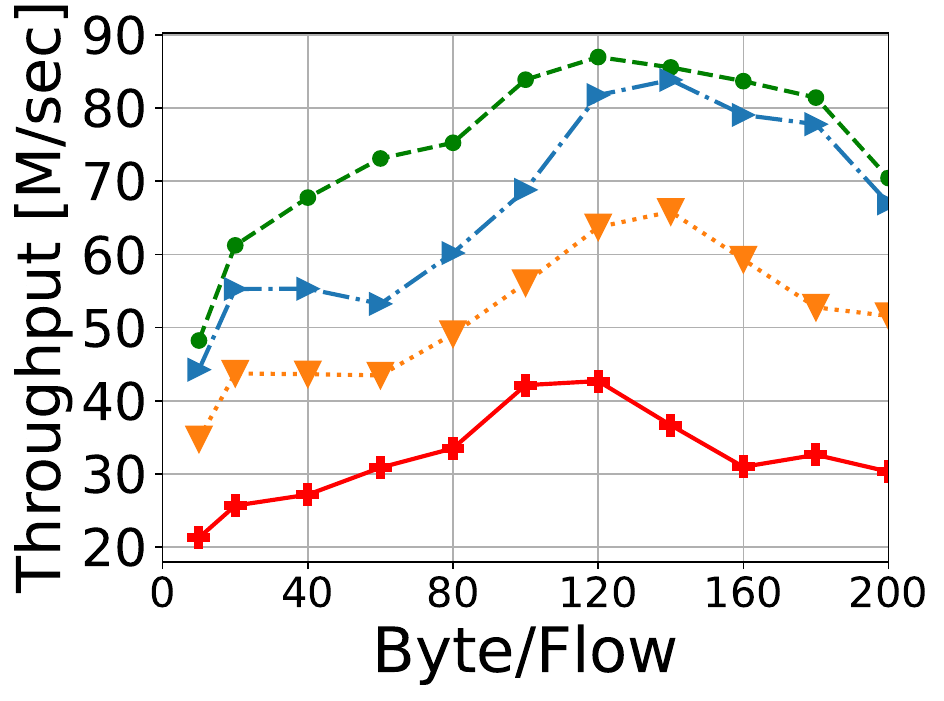}}
        \subfloat[ZIPF 1.4]
        {\includegraphics[width =0.66\columnwidth]    
        {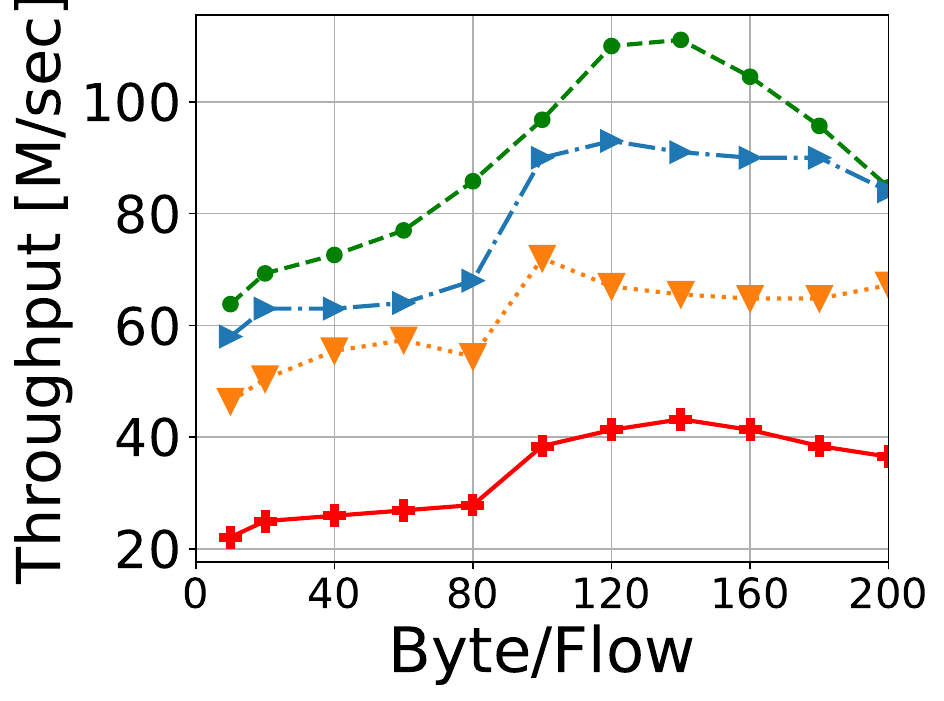}}
        
        \centering
        \caption{Comparing the throughput of counting with hash tables when varying the allocated memory per table entry (flow).\label{fig:cuckoo_speed}}
      \end{figure*} 


\section{Evaluation}
\label{sec:evaluation}

In this section, we evaluate the tradeoffs Counter Pools offer and then compare it to several leading alternatives. 

\noindent\textbf{Datasets:}\label{app:setup}
We used the following datasets to evaluate our algorithm on real network traffic, and on synthetic settings that encapsulate distributions seen in many domains. Specifically, we use the following datasets: 
\begin{enumerate}
\item The CAIDA Anonymized Internet Trace ~\cite{CAIDA} from New York City (Caida18) with $98$M packets.
\item The generated Zipf datasets choosing the Zipf parameter as $0.6$, $1.0$ and $1.4$ (as common in switching caching works~\cite{switchkv, netcache, distcache}). Each dataset has $98$M items.
\end{enumerate}

\subsection{Configuring Counter Pools}\label{app:cp_setup}

We first examine different possible configurations. As we defined Section~\ref{sec:Layout}, a Counter Pool configuration is defined by four numbers $(n, k, s, i)$. Where $n$ is the pool size, $k$ is the number of different counters in the pool, $s$ is the starting size of each pool, and $i$ is the granularity of the element. The writing (64, 4, 0, 1) in the legend refers to $n=64, k=4, s =0,i=1$.

We measure the On Arrival error for count-min-sketches with four rows whose total memory size is in the $x$ axis. The error is measured using Normalized Root Mean Squared Error (NRMSE), which is a standard metric (e.g., see~\cite{SALSA} and the references therein) which is defined as follows. Suppose that the input stream is $x_1,x_2,\ldots,x_n$, and the true frequency of $x_i$ \emph{at time $i$} is $f_i$ and the estimated frequency at that time is $\widehat{f_i}$, then the Mean Squared Error is defined as $MSE=\frac{1}{n}\sum_{i=1}^n(f_i-\widehat{f_i})^2$, and the NRMSE as $NRMSE=\frac{1}{n}\sqrt{MSE}$. Importantly, NRMSE is a normalized metric that is always between 0 (no error) and 1 (no information captured) and thus allows standardized comparison across datasets and measurement lengths.

\vspace{1mm}
As shown in Figure~\ref{fig:conf_cmp}, configurations with four counters per pool yield lower errors than the ones that use more. The reason for that is pool failures. Thus, the desired configuration depends on the average pool size and the chance of failure. 
Figure \ref{fig:conf_cmp_HH} repeats the above experiment but only focuses on the error for the largest heavy hitters. Here, we use the Average Relative Error as the error metric. Using the same notations as before and using $H$ to denote the set of heavy hitters, we have $ARE=\frac{1}{|H|}\sum_{i\in H} \frac{|f_i-\widehat{f_i}|}{f_i}$.
The results here are similar in essence to the previous experiment since most pool failures happen to pools that contain a heavy hitter. 

\vspace{0mm}
{
In both figures, we observe that the (64, 4, 0, 1) configuration is competitive with the other configuration. Such a configuration's practical implementation merits due to memory alignments and fine counter granularity. Thus, we hereafter use this configuration for its flexibility and implementation efficiency. To see why such flexibility is desired, observe Figure~\ref{fig: CounterSizeCmpNYC} that depicts the behavior of real networking traces where the vast majority of counters are small, but there is a small percentage of large counters. Namely, 99\% of the counters require less than 8 bits, making the probability of failure very low. 
}

\subsection{Pool Failure Recovery Mechanisms}
Our next experiment in Figure~\ref{fig: cmpFail} determines the good approach to use upon a pool failure in the sketches use case, where we are allowed to handle failures by a slight increase in error. Our baseline is to not handle pool failures (`Without failing counters'), in which the estimate is based on the non-failing pools (in other sketch rows). Another alternative is offloading failed pools to a secondary data sketch structure  (`X\% failing counters') with fewer counters. Here, the key is the overflowing counter index, which is then hashed into one of the secondary sketch's counters. The last method is to merge the failing pool into two 32-bit counters, allowing the same counting range as baseline sketches.
In this figure, we use the count min sketch and our chosen configuration of (64, 4, 0, 1). The $x$ axis is the size of the sketch, while the $y$ axis is on arrival NRMSE. As can be observed, all methods are roughly the same \mbox{for large sketches since overflows are less common. }

In smaller sketches, we find that the merge approach works better than failure counters. The explanation lies in failing counters requiring extra memory while still allowing for collisions between the heavy hitters. When merging a single pool into two counters, the chances of colliding with heavy hitters are smaller. 

\subsection{Comparison for Approximate Algorithms}
\paragraph{\textbf{Count Min Sketch}}
In Figure~\ref{fig:conf_cmp_alt_HH}, we demonstrate improved heavy hitters accuracy for Counter Pools in its optimal configuration (i.e., if you could choose the configuration in hindsight) and our selected (64, 4, 0, 1) configuration. 
The (64, 4, 0, 1) configuration improves on the alternatives' error for the entire range. The reason for this is that it does not increase the number of hash collisions for the heavy hitters as previous works do~\cite{SALSA,PyramidSketch,gong2017abc}. In essence, we allocate to each counter exactly the number of bits it requires. While the optimal Counter Pools configuration is even more accurate, our goal is to show that a single configuration, which is amenable to efficient implementation and does not require parameter tuning, already offers a more accurate alternative to state-of-the-art solutions.
Notice that for Figures~\ref{fig:conf_cmp_alt_HH} (b) and (f), due to the low skew (0.6), starting from a fairly low threshold, there are no heavy hitters.

Figure \ref{fig:cmp_speed_OnArrival} (a)-(d) show the on-arrival evaluation. Once again, Counter Pools is significantly more accurate throughout and, on many occasions, requires less than a quarter of the memory for the same error levels as the next-best alternative.
Also, as a result of the more sophisticated encoding, as shown in~Figure \ref{fig:cmp_speed_OnArrival}, Counter Pools is about 20\% slower than the baseline while being significantly faster than Abc Sketch. It is also noticeable that for all algorithms, the throughput declines as the size of the sketch is increased. This can be attributed to lower cache residency.

\paragraph{\textbf{Conservative Update Sketch}}
To show the generality of our approach, we evaluate a second sketch, the Conservative Update.
As we show in Figure \ref{fig:cus_speed_OnArrival}, the Counter Pools Conservative Update variant has a lower OnArrival error than the SALSA and Baseline variants, especially for the real-world NYC 2018 trace. However, Counter Pools still exhibit lower throughput than these algorithms. As before, when allocating memory that is larger than the cache size, we observe a performance degradation due to cache misses. In all datasets and memory footprints, SALSA provides a middle ground being slower than the baseline but somewhat more accurate.

\newpage
\subsection{Comparison for Exact Algorithms}
We next compare our Perfect Cuckoo Filter (PCF)-based Counter Pools histogram method to an adaption of the standard PCF that includes values and two off-the-shelf hash maps, including the popular std::unordered\_map and the highly optimized Robin Map~\cite{RobinMap}. 

As Figure~\ref{fig:cuckoo_speed} demonstrates, our Counter Pools-based method offers a speed improvement for computing a histogram. 
The reason for the speedup is the reduced load factor: standard hash maps such as std::unordered\_map and Robin Map allocate 4 bytes for the key and additional 4 bytes for the value. The standard PCF adaptation allows compressing the keys and, when using $2^{17}$ buckets requires just two bytes per key for a total of six bytes per entry.
Finally, our Counter Pools-based PCF further compresses the counter values; using the (64, 4, 0, 1) configuration, we require an average of $20$ bits per counter, for an overall 4.5 bytes per entry. Thus the load factors are, taking the 10 bytes/flow example (leftmost point in the figure): 80\% for the baseline hash maps, 60\% for the PCF adaptation, and 45\% for our Counter Pools PCF. When increasing the memory, up to a certain size, the speed of all algorithms increases due to the lower load, but Counter Pools remain the fastest throughout. As in the sketch experiments, when the size of the data structure exceeds the cache size, we observe a throughput degradation, which implies that exceedingly low load factors are not beneficial.


\section{Discussion}\label{sec:discussion}
In this study, we proposed Counter Pools, a variable-sized counter design that is based on encoding multiple counters within a single memory pool (e.g., 64 bits). This design ensures that, as long as the total bits used by all counters within a pool remain within the pool's capacity, there will be no overflows even if some counters become very large. Counter Pools are especially beneficial for heavy-tailed workloads, such as network traces, where there is a large number of low-frequency items.

For sketch use cases, our findings indicate that Counter Pools improve accuracy across different error metrics and workloads compared to leading alternatives. Our method's primary drawback is a performance slowdown of approximately 20\%, which makes it suitable for applications where accuracy is of the utmost importance.

Additionally, our work extends to exact counting use cases. We demonstrated that counting with a Cuckoo hash table is faster when using Counter Pools than with fixed-sized counters. This is because compressing the counters reduces the table's load factor for the~same~memory usage.

Overall, with about 15 million operations per second for sketches and 35-110 million for hash table manipulations, both applications are practical for most network measurements and accounting use cases with improved accuracy/memory tradeoffs.

\bibliographystyle{IEEEtran}
\bibliography{references}
\end{document}

%% file: macros.tex
\newcommand{\set}[1]{\left\{#1\right\}}

\newcommand{\ceil}[1]{ \left\lceil{#1}\right\rceil}
\newcommand{\floor}[1]{ \left\lfloor{#1}\right\rfloor}

